\documentclass[aps,preprint,superscriptaddress,showpacs,nofootinbib]{revtex4}

\usepackage{latexsym}
\usepackage{amssymb}
\usepackage{epsfig,amsmath,graphics}
\usepackage{pstricks}
\usepackage{color}
\usepackage{dcolumn}

\newcommand{\OO}{\mathcal{O}}

\newcommand{\GeV}{\text{GeV}}
\newcommand{\gev}{\text{GeV}}

\newcommand{\tev}{\text{TeV}}

\newcommand{\tb}{\ensuremath{\tan\beta}}
\newcommand{\vev}[1]{{\langle #1 \rangle}}

\usepackage{comment}
\newcommand{\be}{\begin{equation}}
\newcommand{\ee}{\end{equation}}

\def\bea{\begin{eqnarray}}
\def\eea{\end{eqnarray}}
\def\ltap{\ \raise.3ex\hbox{$<$\kern-.75em\lower1ex\hbox{$\sim$}}\ }
\def\gtap{\ \raise.3ex\hbox{$>$\kern-.75em\lower1ex\hbox{$\sim$}}\ }
\def\lsim{\ \raise.3ex\hbox{$<$\kern-.75em\lower1ex\hbox{$\sim$}}\ }
\def\gsim{\ \raise.3ex\hbox{$>$\kern-.75em\lower1ex\hbox{$\sim$}}\ }

\usepackage{slashed}
\newcommand{\met}{\slashed {E}_{T}}
\usepackage{color}

\begin{document}
\preprint{FERMILAB-PUB-12-388-A-T}

\title{Supersymmetry with a Sister Higgs\footnote{Work Presented at MCTP Symposium on Higgs Physics, April 16-20, 2012}
}

\author{Daniele S. M. Alves}
\affiliation{Center for Particle Astrophysics,\\ Fermi National Accelerator Laboratory, Batavia, IL 60510, USA}
\author{Patrick J. Fox}
\affiliation{Theoretical Physics Department,\\ Fermi National Accelerator Laboratory, Batavia, IL 60510, USA}
\author{Neal Weiner}
\affiliation{Center for Cosmology and Particle Physics, \\Department of Physics, New York University, New York, NY 10003}
\date{\today}
\begin{abstract}
Within the context of supersymmetric theories, explaining a 125 GeV Higgs motivates a consideration of a broader range of models. We consider a simple addition to the MSSM of a ``Sister Higgs'' ($\Sigma_d$), a Higgs field that participates in electroweak symmetry breaking but does not give any direct masses to Standard Model matter fields. While a relatively minor addition, the phenomenological implications can be important. Such a field can be naturally charged under an additional symmetry group $G_s$. If gauged, the Higgs mass is naturally much larger than in the MSSM through an NMSSM-type interaction, but with $\Sigma_d$ playing the role of $H_d$. The addition of the sister Higgs allows new R-parity violating operators $\Sigma_d H_d E$, which are less constrained than conventional leptonic R-parity violation. Considerations of unification motivates the presence of colored $G_s$-charged fields. Production of these G-quarks can lead to new b-rich final states and modifications to decays of gluinos, as well as new opportunities for R-parity violation. Unlike a conventional fourth generation, G-quarks dominantly decay into a light jet and a scalar (potentially the Higgs), which then generally decays to b-jets. The presence of additional sister charges allows the possibilities that lightest sister-charged particle (LSiP) could be stable.  We consider the possibility of an LSiP dark matter candidate and find it is generally very constrained. \end{abstract}

\maketitle

\section{Introduction}

As data continue to arrive from the LHC, the simplest models of supersymmetry have become increasingly constrained. In particular, the recent discovery of a Higgs-like feature at 125 \gev~\cite{cmshiggstalk,atlashiggstalk} is tantalizingly close to the tree level prediction of the MSSM, but is large enough to require large radiative corrections, and pushes the MSSM into narrow corners of parameter space \cite{Hall:2011aa}. At the same time, no clear evidence for any of the other superpartners has appeared.  Both of these results can be accommodated by a higher SUSY scale, but at the cost of increasing the tuning of electroweak symmetry breaking scale.

Alternatives for raising the Higgs mass have been varied:  the simplest possibility maybe the NMSSM, in which a singlet $S$ is added to the MSSM together with the superpotential term $\lambda S H_u H_d$. The resulting quartic coupling $|h_u h_d|^2$ raises the Higgs mass at small $\tb$. However, aside from the aesthetic issues with including a complete singlet, achieving a large enough quartic requires a superpotential coupling which must be near the limit of perturbativity at the GUT scale. At the same time, if a much larger value for the quartic were possible, then improved possibilities for naturalness arise \cite{Hall:2011aa}.

Large quartics can be natural, for instance if the singlet $S$ and/or Higgs is a composite \cite{Harnik:2003rs,Chang:2004db} (alternatively the new states can be integrated out yielding quartics from K\"ahler potential terms \cite{Evans:2012uf}). New matter at high energies can boost the quartic somewhat by raising the standard model gauge couplings in the UV \cite{Espinosa:1992hp,Masip:1998jc}. 

While the D-term quartics are fixed by supersymmetry, they can be enhanced by non-decoupling D-terms from an additional gauge group \cite{Batra:2003nj,Batra:2004vc}. While appealing in principle, most models require significant additional matter content to maintain unification.

Extending the MSSM to include new fields is dangerous however, as they can introduce new sources of FCNCs or significant corrections to precision electroweak observables. There is a simple addition, however, that can evade these concerns: namely, the inclusion of additional Higgs multiplets, which we denote $\Sigma_u$ and $\Sigma_d$.  These fields carry the same SM quantum numbers as $H_u$ and $H_d$, respectively. To avoid FCNCs these fields should have no tree-level, renormalizeable couplings to SM fermions. At the same time, there is no reason that they cannot participate in EWSB by acquiring significant vevs. Such fields we refer to as ``sister'' Higgs fields.

The inclusion of additional Higgs fields is certainly not new. What we are describing is largely the generalization of the Type I 2HDM to a supersymmetric context. Similar non-coupling Higgses were employed to construct R-symmetric SUSY theories \cite{Kribs:2007ac}. As we shall explore in this paper, however, there are important phenomenological consequences when allowing fields such as these in the low energy theory. 

First, an important point is that for NMSSM-like enhancements of the Higgs mass, the coupling $S H_u \Sigma_d$ is just as effective as $SH_u H_d$, i.e., the object that acquires a large vev needn't be the object that gives the down-type fermions their masses. This allows a great deal of freedom in reconsidering the dynamics of the light fields in the theory (i.e., the charged Higgs) and whether they carry additional quantum numbers under some new group $G_s$.  In particular, the charges of $\Sigma_d$ are only tied to the charges of S, not to the charges of down-type fermions.

Secondly, the inclusion of the sister Higgs allows the presence of a new R-parity violating operator, namely $H_d \Sigma_d E$, and with it comes a great deal of new phenomenology. In the presence of R-parity conservation, new LSPs are possible, including LSPs that carry a conserved charge and thus can potentially provide a dark matter (DM) candidate.

Finally, the presence of these fields motivates the presence of additional down-type quarks, also charged under $G_s$. These G-quarks can have important consequences for signals such as gluino decays, and other opportunities for R-parity violation. Unlike usual 4th generations, these G-quarks decay via scalar (often Higgs) emission, leading to b-rich final states.

Any one of these points motivates a consideration of these theories with extended Higgs sectors.  We describe the construction of a sister MSSM in more detail below.  In the rest of the paper we discuss the extension to a gauged sister model, before addressing each of the above mentioned points in more detail.

\subsection{A Sister Higgs}\label{sec:sh}

A sister Higgs is a Higgs field that participates in EWSB but has no tree level couplings to matter. 
This is a natural outcome when the sister Higgs transforms as a non-trivial representation of some new symmetry $G_s$ under which the SM fields are trivial.
This may be a gauged or global symmetry. As we shall see shortly, although in its simplest manifestation $G_s$ is not gauged, it is likely most interesting in a case when $G_s$ is a gauge symmetry.

\begin{table}[t]
\centering
\begin{tabular}{c|c|c} 
 & $SU(3)\times SU(2)\times U(1)_Y$ & $G_s$ $[SU(2)_s]$ \\ 
 \hline  
$\Sigma_u$     &  $(1,2,1/2)$                         &  $\mathbf{r}$ $[2]$ \\
$D_g$             &  $(\mathbf{3},1,-1/3)$          &  $\mathbf{r}$ $[2]$ \\
$\Sigma_d$     &  $(1,2,-1/2)$                        &  $\mathbf{\bar{r}}$ $[2]$ \\
$D_g^c$          &  $(\bar{\mathbf{3}},1,1/3)$  &  $\mathbf{\bar{r}}$ $[2]$ \\
$\Phi$              &  $(1,1,0)$                            &  $\mathbf{r}$ $[2]$ \\
$\bar{\Phi}$      &  $(1,1,0)$                            & $\mathbf{\bar{r}}$ $[2]$\\
$H_u$              &  $(1,2,1/2)$                         &  $1$ $[1]$\\
$H_d$              &  $(1,2,-1/2)$                         &  $1$ $[1]$\\
\end{tabular}
\caption{Charges of fields under the SM and the Sister group, $G_s$, for the generic case [the case of $G_s=SU(2)$].  Together $\Sigma_u$ and $D_g$ ($\Sigma_d$ and $D_g^c$) make a $\mathbf{5}$ ($\mathbf{\bar{5}}$) under $SU(5)$. Note the usual MSSM fields $Q,U^c,D^c,L$, and $E^c$ are singlets under $G_s$.}
\label{tab:charges}
\end{table}

To make this more concrete: we extend the MSSM by introducing two sister Higgs fields, $\Sigma_u$ and $\Sigma_d$, which carry the same charge under the SM as $H_u$ and $H_d$ respectively, while also being vector-like under the new sister group, $G_s$.  At this stage we do not specify whether $G_s$ is a global or gauged symmetry.  In order to allow couplings between the sister fields and the MSSM we also introduce fields that only carry charge under $G_s$.  Furthermore, if these fields acquire a non-zero vacuum expectation value (vev) then there will be mass mixing between the sister Higgses and the Higgses of the MSSM.  We denote these $G_s$ breaking fields as $\Phi (\bar \Phi)$, which are in the same representation of $G_s$ as $\Sigma_u (\Sigma_d)$.  In addition to these fields, as we will see, it is natural to expect vector-like fields that transform as right-handed quarks and are also charged under $G_s$, which we dub G-quarks, $D_g, D_g^c$.  Later, in Section~\ref{sec:higgsmass}, we will investigate a particularly interesting case where the sister group is a gauged $SU(2)$.  The charges of all fields in both the general and this special case are shown in Table~\ref{tab:charges}.  With this field content and assuming ${\bf r} \ne {\bf \bar r}$ the simplest extension of the MSSM superpotential is 
\bea
W &=& \mu_\phi \Phi \bar \Phi+\mu_\Sigma \Sigma_u \Sigma_d +\mu_h H_u H_d+ \mu_g \bar{D}_g D_g 
\nonumber \\
&&+\lambda_u \Phi H_u \Sigma_d + \lambda_d \bar \Phi \Sigma_u H_d \nonumber\\
&& + Y_u U^c H_u Q + Y_d D^c H_d Q + Y_\ell E^c H_d L ~.
\label{eq:SHsup}
\eea
To make this completely general, we can add a term $\bar \Phi D_g D^c$ that we shall discuss shortly. Notice that if the representation $\mathbf{r}$ is different from the conjugate $\mathbf{\bar{r}}$, this superpotential contains an additional $U(1)$ symmetry, which needn't be accidental, which we denote $U(1)_{\Sigma}$. Under this symmetry $\Sigma_u,  \Phi$ have charges $1/2$ and $\Sigma_d, \bar \Phi$ have charges $-1/2$.   In the interesting case where $G_s=SU(2)$, $\mathbf{r}$ is pseudo-real, and operators $\Phi \Sigma_u H_d+ \bar \Phi \Sigma_d H_u$ are also allowed, for instance, which break the $U(1)_\Sigma$.
 Although the concept of a Sister Higgs is independent of $G_s$, if this group is too large (if the representations of $\Sigma$ are larger than 4 dimensional) then the SM gauge couplings will reach a Landau pole below the GUT scale.

The most natural origin for SM fermion masses is through the usual Yukawa couplings to $H_u$ and $H_d$ as it is in the MSSM, and we have illustrated this in (\ref{eq:SHsup}). However, with the inclusion of new fields acquiring vevs, it is possible that down-type fermion masses can instead come from the higher dimension operator $\Phi\Sigma_d Q D^c$.  This could be generated by integrating out a massive ${\bf 5 + \bar 5}$ pair, which are uncharged under $G_s$, and have couplings $\mathbf{\bar{5}}QD^c + \mathbf{5}\Phi \Sigma_d$. We will not focus on this case, but it may be natural to consider extending this model by the presence of additional $\mathbf{{5}}+\mathbf{\bar{5}}$'s.

\section{Gauged Sisters and the Higgs Mass}\label{sec:higgsmass}
As we have discussed, there has been great activity attempting to accommodate a Higgs at 125 GeV in supersymmetric theories. Such a large Higgs mass in the MSSM requires squark masses in the multi-TeV range and large A-terms, well beyond our expectations from naturalness \cite{Giudice:2011cg}.

Various extensions of the MSSM attempt to alleviate this problem, as we have discussed.  While the simplest possibility may simply be of a tuning of the MSSM, the simplest potentially natural model is likely the NMSSM. As is well known \cite{Masip:1998jc,Espinosa:1998re}, the RG flow of the $S H_u H_d$ coupling in the NMSSM tends to make it run quite small in the IR. Unlike the top quark, which has $SU(3)$ color contributing to its anomalous dimension, there is no strong interaction to drive this coupling large. Consequently, a 125 GeV Higgs lies near the edge of perturbativity without appealing to sizable radiative corrections from top loops.  This leads us naturally to the possibility of gauging the sister group.  If the sister group runs strong then the $\lambda_u$ and $\lambda_d$ couplings, which play the same role as $S H_u H_d$ in the NMSSM, can remain large and give a sizeable correction to the Higgs self coupling and therefore the Higgs mass.  However, such an approach has a ``why-now'' problem: the coincidence of the  strong coupling scale of the new gauge group and the weak scale.  If instead the running of the gauge coupling of the sister group, $g_S$, were approximately conformal this why-now problem would be removed.  An example of such a setup -- where the beta function is zero at one loop -- is $G_s=SU(2)_s$ with\footnote{This group is also small enough that the additional matter charged under the SM does not lead to Landau poles in the SM gauge couplings.} six vector-like families of matter.  

With $G_s=SU(2)_s$ and $\Phi,\, \bar{\Phi},\, \Sigma_u,\, \Sigma_d$ all in fundamental representations of $SU(2)_s$ a one-loop conformal beta function occurs if there are 3 more $(\square,\overline{\square})$ pairs, the G-quarks.  This is exactly the field content shown in Table~\ref{tab:charges} and is the same as what would be expected if the sister Higgses were embedded in a $(\mathbf{5},\overline{\mathbf{5}})$ of $SU(5)$.  Remarkably, simply by insisting on conformality as a solution to the ``why-now'' problem, we are naturally pushed into a unified (i.e. GUT) setup with the G-quarks. 
While the gauge coupling for $SU(2)_s$ is conformal at one-loop, the SM values are modified by the addition of two additional flavors, i.e., $\beta_3 = -1, \beta_2 = 3, \beta_1 = 43/5$.

The low energy phenomenology of the sister Higgs scenario can be much more complicated than the usual MSSM (or NMSSM). In general, there are as many as six ``Higgs'' fields (i.e., $SU(2)_W\otimes U(1)_Y$ doublets) as well as four SM singlet scalars, in addition to a $Z'$ and the associated fermionic fields.  However, for the most part, we can understand the phenomenology by taking a simplified limit of the theory. As we will see this limit reduces the theory to a Type I 2HDM \cite{Barger:1989fj}. \footnote{In a Type I 2HDM, there is a standard model like Higgs $H_1$ which couples to fermions and a second Higgs $H_2$ which does not, although it can acquire a vev. A mixing between them induces couplings for both mass eigenstates.} 

We can take this limit as follows: let us begin by considering the superpotential in (\ref{eq:SHsup}).  This superpotential retains a global $U(1)_s=U(1)_\Sigma+T^3_s$ symmetry after $SU(2)_s$ breaking and an associated ``sister charge".  Within the $SU(2)_s$ doublet $\Sigma_d$, for instance, only one component mixes with the Higgs, while the other component, carrying the sister charge, will not. When writing the scalar potential, we will designate $\Sigma_d^s$ with the $^s$ to indicate when the field carries sister charge, and similarly with the $\Phi$, $\bar \Phi$ and $\Sigma_u$. In discussing the Higgs sector, we will for the moment assume the presence of this symmetry as it simplifies the physics by reducing by a factor of two the number of fields that mix with the Higgs boson.

Since $\Sigma_d$ will take the role of the $H_d$ for the purposes of raising the Higgs mass, $\lambda_d$ is not important for the Higgs mass phenomenology, so for simplicity we set it to zero.  We take $SU(2)_s$ to be broken by a $\vev{\bar \Phi} = v_\phi \cos \beta_s$ and $\vev{\Phi} = v_\phi \sin \beta_s$. In general, we will be taking $\tan \beta_s \ll 1$, so the $\bar \phi$ is acquiring the larger vev. For the moment, we then set $\beta_s=0$.
We will assume that the $\phi$ fields have large enough SUSY breaking masses that they can be integrated out without disturbing the phenomenology (i.e., the mass spectrum and the $\lambda_u^2 |H_u \Sigma_d|^2$ quartic) and set them to their vevs. $\Sigma_u$ is for the most part a spectator, so we integrate it out with a SUSY breaking mass. 

The resulting theory, with these assumptions has a vev-acquiring sector composed of $\Sigma_d$, $H_d$ and $H_u$, with the scalar potential,
\bea
V &= &m_{H_d}^2 |H_d|^2 + m_{H_u}^2 |H_u|^2 +(m_{\Sigma_d}^2+\frac{m_{Z_s}^2 }{2})| \Sigma_d|^2+(m_{\Sigma_d}^2-\frac{m_{Z_s}^2}{2})| \Sigma^s_d|^2\\
&&+ |\lambda_u|^2 |H_u \Sigma_d|^2 +(B \mu H_u H_d+ \lambda_u \mu_\phi^* \langle\bar \Phi\rangle^* H_u \Sigma_d + {\rm h.c.})+\frac{1}{2}D_{MSSM}^2+\frac{1}{2}D_{s}^2~.\nonumber
\eea
Here, we have taken the fermions to couple to the scalar sector through the usual Yukawa matrices $Y_u Q U^c H_u$ and $Y_d Q D^c H_d$ and $D_{MSSM}$ are the usual D-term contributions from $SU(2)_W$ and $U(1)_Y$, and $D_s$ are the residual $SU(2)_s$ D-term quartics after integrating out the $\phi, \bar \phi$. 

We can further integrate out $H_d$ in the large $m_{Hd}^2$ limit by $H_d = B \mu H_u^*/m_{Hd}^2$, which yields the final simplified setup. 
\bea
V &=  & m_{H_u}^2 |H_u|^2 +(m_{\Sigma_d}^2+\frac{m_{Z_s}^2 }{2})| \Sigma_d|^2+(m_{\Sigma_d}^2-\frac{m_{Z_s}^2}{2})| \Sigma^s_d|^2 \\
&&+ |\lambda_u|^2 |H_u \Sigma_d|^2+ (\lambda_u \mu_\phi^* \langle\bar \Phi\rangle^* H_u \Sigma_d + {\rm h.c.}) +\frac{1}{2}D_{MSSM}^2+\frac{1}{2}D_{s}^2~,\nonumber
\eea
where the fermion masses now arise in the effective theory via the terms $Y_u H_u Q U^c $ and $Y_d \frac{B\mu}{m_{Hd}^2} H_u^* Q D^c $. Put simply, the low energy effective theory is a Type-I 2HDM with an additional spectator $\Sigma_d^s$. Although $\Sigma_d^s$ does not mix with these fields nor acquires a vev, we have left it in because of the relationship between its mass and that of the $U(1)_s$ neutral $\Sigma_d$ (which can lead to tensions in some regions of parameter space).  From here, the scalar phenomenology is fairly simple. $H_u$ acquires a vev $\vev{H_u} = v \sin \beta$, while $\vev{\Sigma_d} = v \cos \beta$, and the Higgs quartic has an NMSSM like contribution proportional to $\lambda_u^2$.

The NMSSM-like boost to the Higgs mass will depend on how large $\lambda_u$ can be.  The $\beta$ function for $\lambda_u$ at one loop is
\be
\beta_{\lambda_u}=\lambda_u \left( 3 y_t^2+5 \lambda_u^2-\frac{3}{5} g_1^2 - 3 g_2^2-3g_s^2\right),
\ee
where $g_{1,2,3}$ are the standard model couplings, and $g_s$ is the coupling of the new sister group.  These expressions make clear the simple point: the presence of $g_s$ acts to counter the tendency of $\lambda_u$ (and similarly for $\lambda_d$) to run small in the IR.  A simple question, therefore, is how large can the Higgs mass be in this theory, consistent with perturbativity up to the GUT scale?  Because some couplings are sizeable, it is best to calculate the RG flow to low energies at two-loop.  To do this, we use the publicly available \texttt{SARAH} package \cite{Staub:2008uz,Staub:2010jh}, to evolve the couplings within our theory from $M_{GUT}=2\times 10^{16}\,\gev$ to 1 TeV.  We generate a large set of random high scale values of $g_s,\, y_t,\,\lambda_u$, all lie in the range $[0,4]$, and run down to low energies. We define $\sin \beta =\frac{ m_t(1\,\tev)}{ v~ y_t(1\,\tev) }$, where $v\equiv 174~\GeV$ and we take $m_t(1\,\tev)=150.7\,\gev$~\cite{Xing:2007fb}. Points with $\sin \beta >1$ are disregarded as in conflict with observation.

\begin{figure}[t] 
   \centering
   \includegraphics[width=0.7\textwidth]{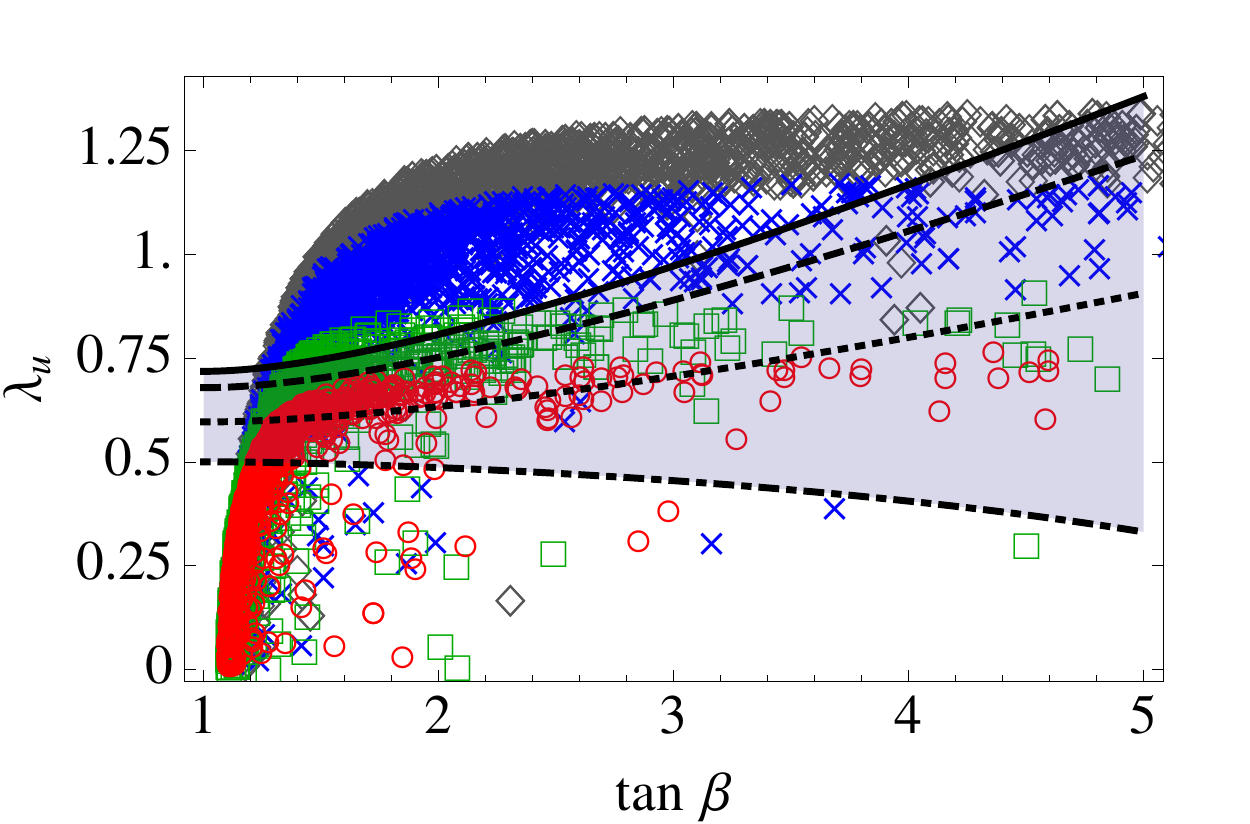} 
   \caption{IR values of $\tan\beta$ and $\lambda_u$ when scanning over the range [0,4] for the UV inputs $g_S$, $\lambda_u$, and $y_t$.  Different ranges for $g_S(1\,\tev)$ are denoted by different symbols, $\bf\red{\circ}$: $g_S<0.5$, $\bf\green{\Box}$: $0.5\le g_S <1$, $\bf\blue{\times}$: $1\le g_S<1.5$, and $\bf\gray{\Diamond}$: $g_S\ge 1.5$.  The lines show the parameters for which a higgs of 125 GeV is achieved at tree level (solid), with stops of 250 GeV (dashed), 500 GeV (dotted) and 1 TeV (dot-dashed).}
   \label{fig:tanbetalambda}
\end{figure}

The results of our scan are shown in Figure \ref{fig:tanbetalambda}. While one must be careful with this plot (as there is no sense of a measure on this scatter plot) we can get a sense of correlations between parameters.  First, we see that it is quite easy to have very large values of $\lambda_u$ at low energies. Taking $\alpha_s(M_{GUT}) \gsim 0.18$, i.e., $g_s(M_{GUT})> 1.5$, we can have $\lambda_u \simeq 1.25$ at low energies. This is sufficiently large that no additional correction from stop loops is necessary.  However, including even moderate stop contributions is easy. With $\alpha_s = .02$ (i.e., comparable to the $SU(2)_W$ coupling) $\lambda_u$ is easily large enough to get $m_H = 125\, \gev$ with only $\sim 500\, \gev$ stops. 

A final point on which we make a small comment, is that $\Phi$ plays the role of the usual $S$ in the NMSSM and, as recently pointed out by \cite{Hall:2011aa}, for large $\lambda_u$ the theory can become more natural, in that the soft Higgs mass can be larger, and then mixing the SM singlet scalar (in this case $\phi$) with the Higgs, its physical mass can be lowered at the expense of just a moderate tuning. Although we do not explore this quantitatively here, it is clear that the ingredients (a large $\lambda_u$ and a neutral scalar mixing with the Higgs) are present.

\begin{figure}[t] 
   \centering
   \includegraphics[width=0.6\textwidth]{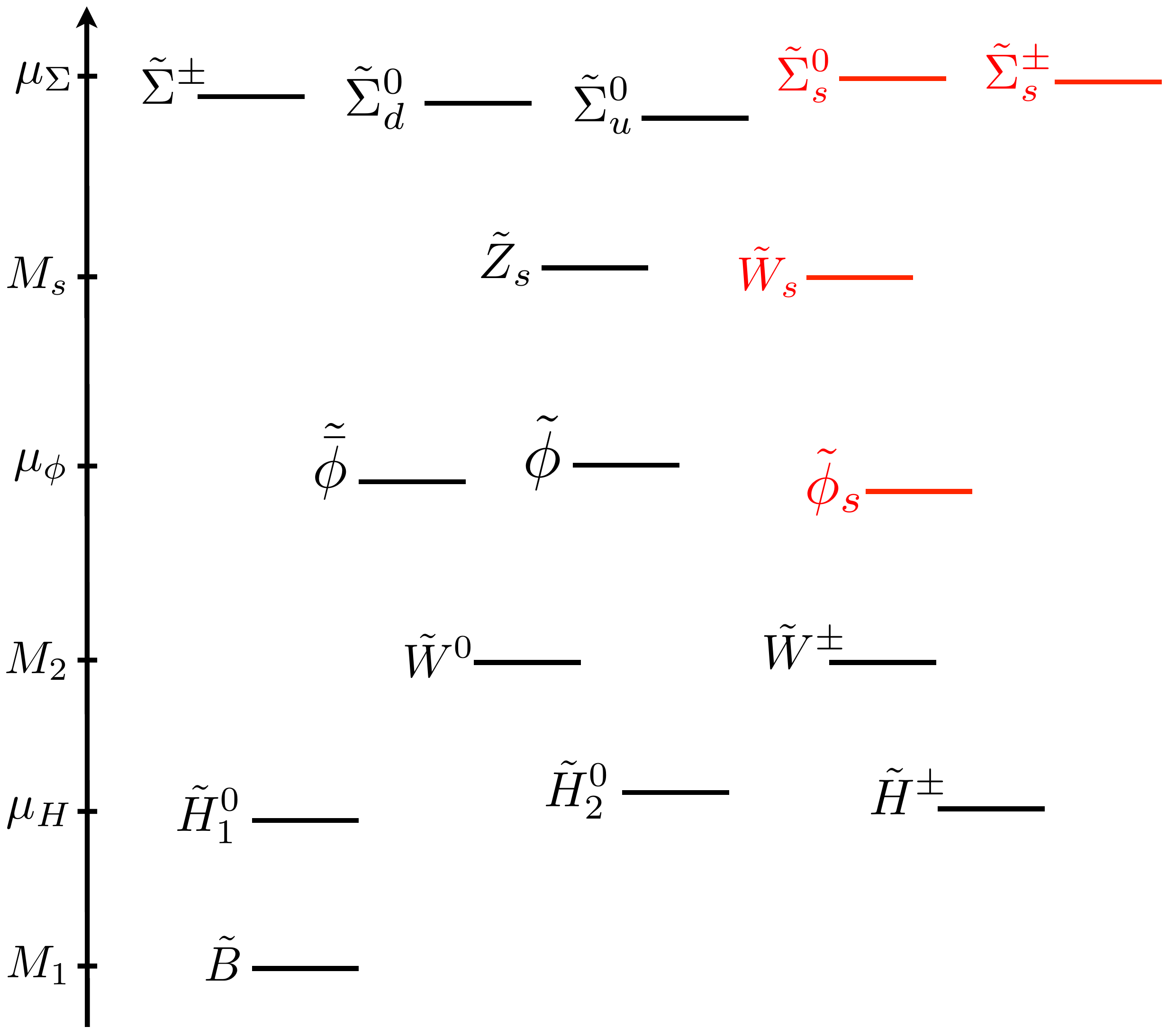} 
   \caption{Example spectrum of fermions, in our simplified limit. States in red are charged under the global $U(1)_s$. No special ordering of mass scales is implied.}
   \label{fig:Fspectrum}
\end{figure}

\begin{figure}[t] 
   \centering
   \includegraphics[width=0.9\textwidth]{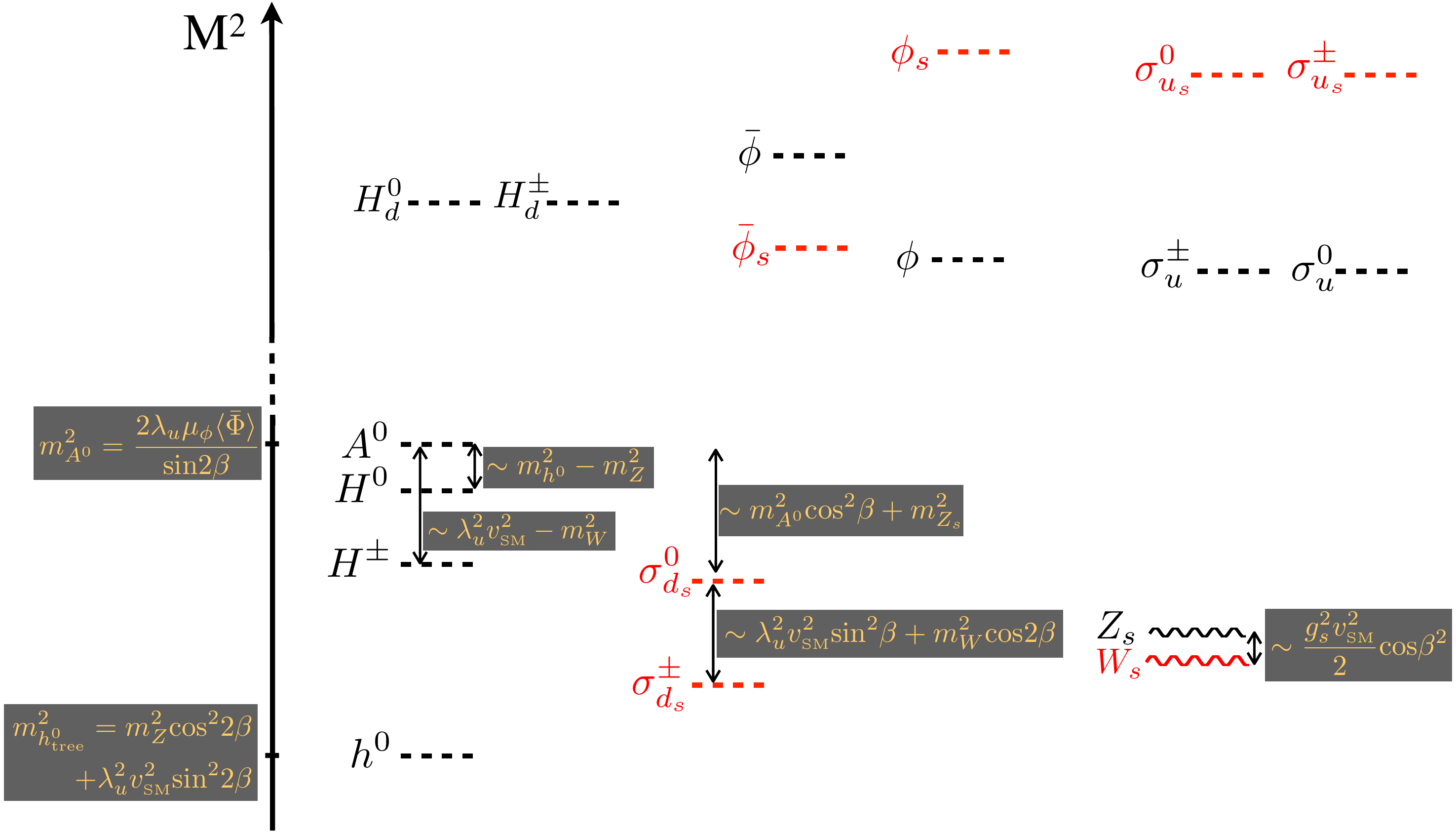} 
   \caption{Example spectrum of scalars and vector bosons, in our simplified limit. States in red are charged under the global $U(1)_s$.}
   \label{fig:Sspectrum}
\end{figure}

\subsection{New Phenomenology in the Simplified Limit}

Even in this limit where we have decoupled $\phi, \bar \phi$ and $\Sigma_u$, there are very important differences from the MSSM.
First, because the second Higgs field in the low energy theory, $\Sigma_d$, has no direct couplings to SM fermions, the mass constraints that apply to $H^\pm$ in Type-II 2HDM do not directly apply. In particular, the charged Higgs can be $\lsim 150 \gev$ without necessarily contributing too largely to the decay of the top, nor excessively to $b\rightarrow s \gamma$ \cite{Mahmoudi:2009zx}.

However, unlike a pure Type-I 2HDM, with $G_s=SU(2)$, the sister Higgs is tied to a second field, namely its sister-charged partner $\Sigma_d^s$. In particular, the mass of the charged Higgs $H^\pm$ (which is dominantly $\Sigma_d^\pm$) is related to the mass of the sister-charged $\Sigma_d^{s^\pm}$ by $m^2_{H^\pm}-m^2_{\Sigma_d^{s^\pm}}\sim m_{Z_s}^2$ (assuming $\vev{\Phi}\ll \vev{\bar\Phi}$). With larger $\vev{\Phi}$, this becomes $m^2_{H^\pm}-m^2_{\Sigma_d^{s^\pm}}\sim  (\cos2\beta_s+2\lambda_u^2\sin^2 \beta_s/ g_s^2)~m_{Z_s}^2$ (assuming $\tan\beta\gg 1$). Either way, for heavy enough $Z_s$ and light enough $H^\pm$, this will induce a charge-breaking negative squared mass to $\Sigma_d^{s^\pm}$.

Probably the simplest way out is to break the global sister charge $U(1)_s$ by including an operator such as $\lambda_u^{\mbox{$s\hspace{-0.145in}\not\hspace{0.1in}$}}\bar \Phi H_u \Sigma_d$. The F-term for $H_u$ then gives an additional contribution to the mass of $\Sigma_d^{s}$, avoiding the unwanted charge-breaking negative $m^2_{\Sigma_d^{s^\pm}}$.  This operator, combined with $\lambda_u\Phi H_u \Sigma_d$, will also contribute to the squared-mass of $H_u$ through the F-term for $\Sigma_d$. In order to avoid fine-tuning, $\vev{\Phi}$ and $\vev{\bar\Phi}$ should not be too large, implying that $Z_s$ should not be too heavy relative to the EW scale. Moreover, including this sister-charge breaking term will affect the running of $\lambda_u$, but given that large values of $\lambda_u$ are already generally quite natural, this is unlikely to be problematic.

\begin{figure}[t] 
   \centering
   \includegraphics[width=0.9\textwidth]{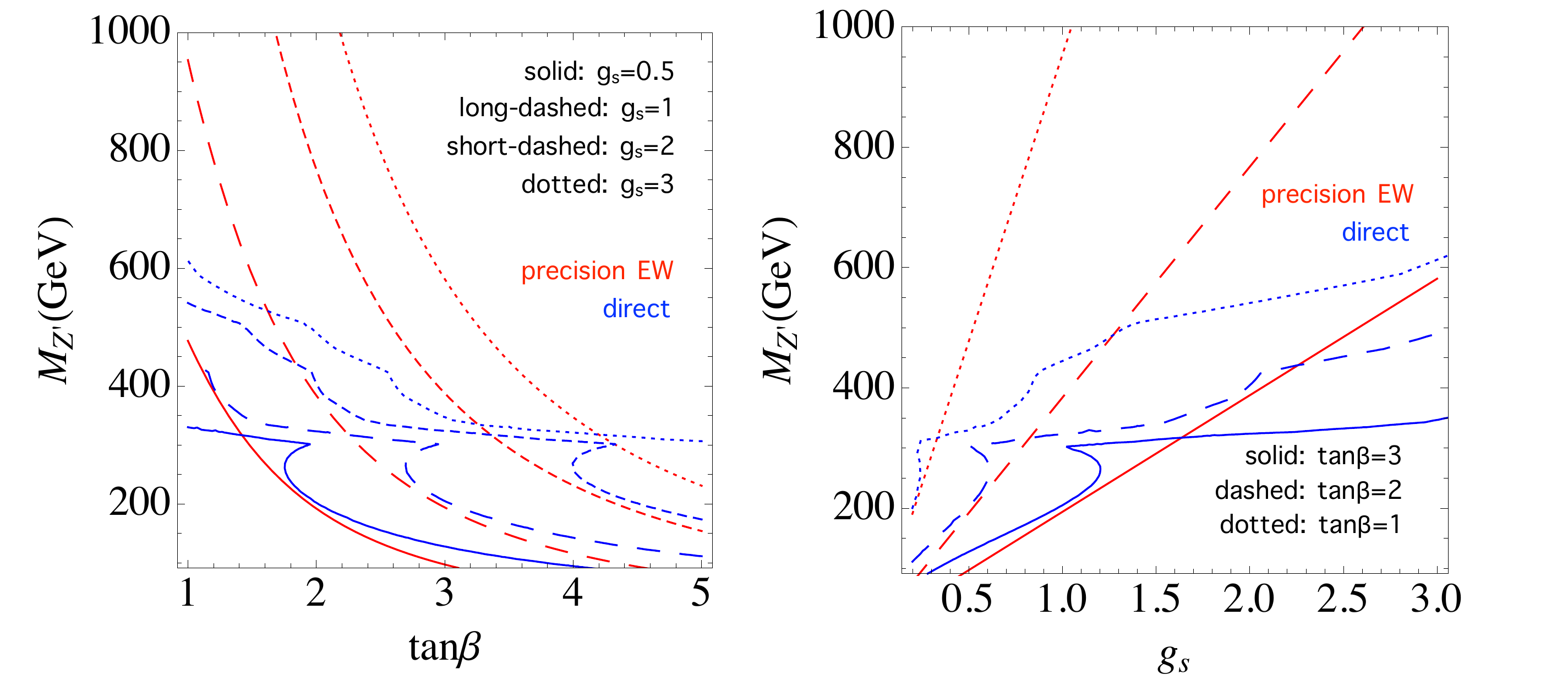} 
   \caption{Bounds on $Z_s-Z$ mixing expressed in terms of $m_{Z_s}$, $\tan\beta$ and $g_s$, from precision electroweak constraints of the $\rho$-parameter (red) and direct collider searches for dilepton resonances \cite{ATLAS-CONF-2012-007} (blue). The later are derived assuming that $Z_s$ has the same branching ratios to leptons as the SM $Z^0$. The region below each line is excluded.}
   \label{fig:Zprimebounds}
\end{figure}

Another key element of this setup is the presence of new gauge forces, $Z^0_s$ and $W^0_s$. Because $\Sigma_d$ is charged under both $SU(2)_s$ and $SU(2)_\text{W}$, its vev will induce mixing between $Z^0_s$ and the SM $Z^0$. Such mixing is constrained by precision electroweak (PEW) measurements of the $\rho$-parameter (although, as has been pointed out previously, e.g., by \cite{Fan:2011vw}, this can actually help agreement with the standard model) as well as collider searches for $Z^\prime$ resonances. ATLAS performed a search for high-mass dilepton resonances in \cite{ATLAS-CONF-2012-007}. We can conservatively re-cast the results of that search in terms of $M_{Z_s}$, $g_s$ and $\tan\beta$, assuming that $Z_s$ decays as a sequential $Z^\prime$ ({\it i.e.} assuming that its branching ratios to SM fermions is identical to that of the SM $Z^0$). These bounds, as well as the PEW constraints, are shown in Fig.\ref{fig:Zprimebounds}. Note that even for large $g_s$ and moderate $\tan\beta$, $Z_s$ can be relatively light. Finally, we point out that the direct bounds from dilepton resonances are significantly weakened if the dominant decay mode of $Z_s$ is $Z_s\rightarrow Z^0 h^0, W^\pm H^\mp$ or even $\phi^*\phi$, which is naturally the case if such on-shell final states are kinematically allowed and $\tan\beta$ is not too large. Despite this, the PEW constraints are stronger and hold regardless of how $Z_s$ decays.

\begin{figure}
   \centering
   \includegraphics[width=0.6\textwidth]{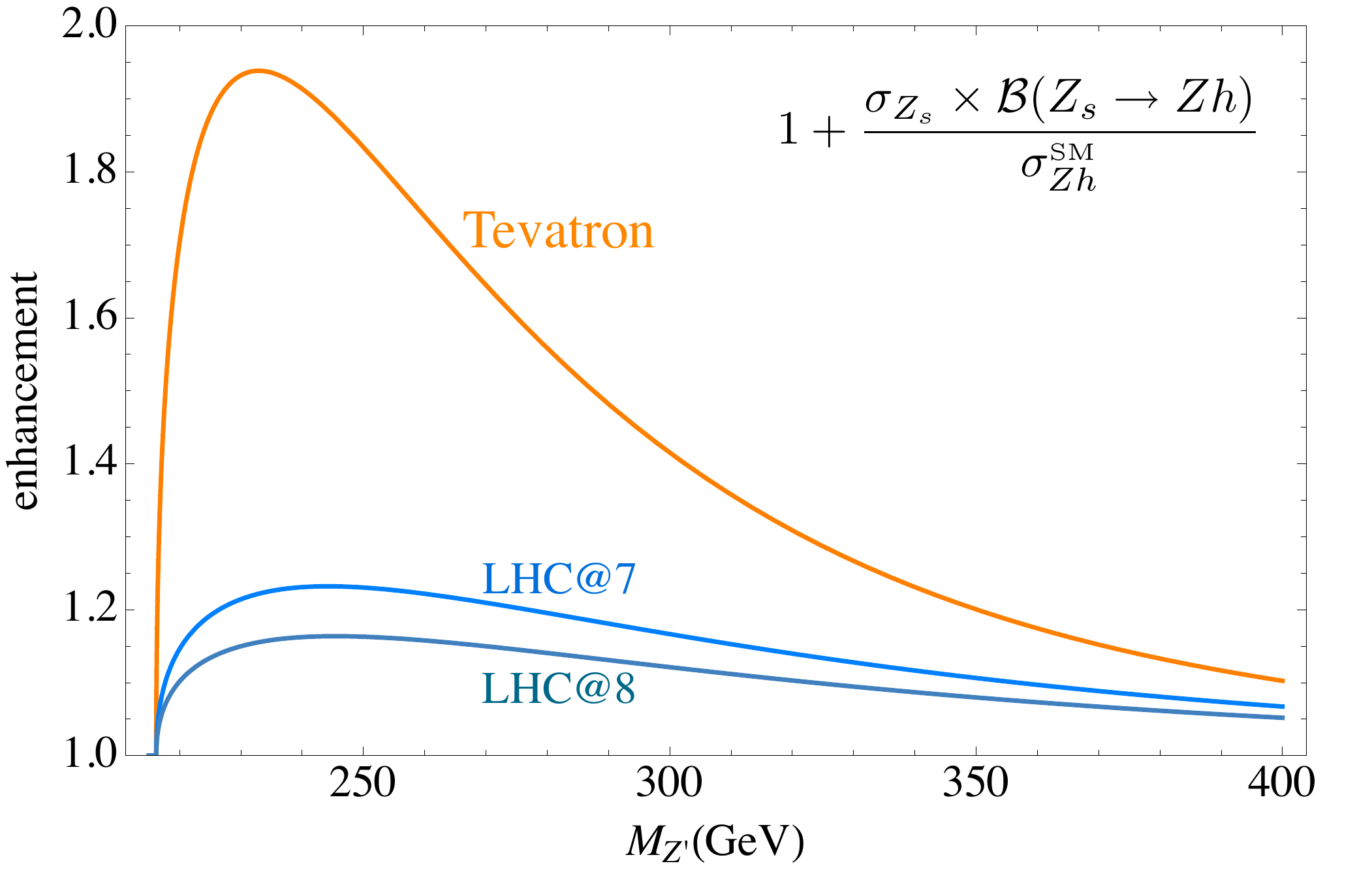} 
   \caption{Saturating the bounds for the $\rho$-parameter, there can be a sizeable contribution to $Z^0h^0$ production at the Tevatron through resonant $Z_s$ production. At the LHC, it is a small correction to the SM process.}
   \label{fig:ZHproduction}
\end{figure}

An interesting consequence of $Z_s\rightarrow Z^0 h^0$ being the dominant decay mode of $Z_s$ is that it provides a sizable contribution to $Z^0h^0$ production at the Tevatron. Saturating the precision electroweak constraint on $Z-Z_s$ mixing, the $Z_s$ production can be comparable to that of SM $Z^0 h^0$ production at the Tevatron for relatively light $Z_s$, as shown in Fig.\ref{fig:ZHproduction}. At the LHC such effect is much smaller, however.

Finally, we comment on the implications of the Sister Higgs framework on the phenomenology of dark matter. As was pointed out in \cite{ArkaniHamed:2006mb}, in the phenomenologically viable region of MSSM parameter space, achieving the correct relic abundance for LSP dark matter requires considerably tuning between parameters.  There are several viable regions where the so-called ``well tempered" neutralino lives, with just the right admixture of bino/wino or bino/higgsino to allow the LSP to be light ($m_\chi\sim\mathcal{O}(100)\,\gev$) and have the correct relic abundance.  In the case of the Sister MSSM there are many new neutralino states that can mix with those of the MSSM; in general the neutralino mass matrix is now $13\times 13$.  Many of these new states carry charge under $G_s$ and have new interactions, either through the superpotential or, if $G_s$ is gauged, $Z_s/W_s$ exchange.  The new gauge bosons can be light and have sizeable gauge couplings (Fig.~\ref{fig:Zprimebounds}), and so have considerable effect on the process of thermal DM freezeout.
The possibility of Dirac dark matter charged under the residual sister $U(1)_s$ will be explored at length shortly, our point here is simply that the neutralino sector is much richer, which allows for a broad range of possibilities for DM.

\section{R-Parity Violation and a Sister Higgs}\label{sec:rpv}

With the addition of the new fields charged under $G_s$, the symmetry structure quickly grows complex. As a consequence, some of the new fields may be stable. At the same time, we shall see that an intriguing new opportunity for R-parity violation appears in this theory.  Let us begin by considering {\em only} the superpotential of eq.(\ref{eq:SHsup}).  This superpotential has the gauge symmetry group $SM \otimes SU(2)_s$, but has a global symmetry $U(1)_B \otimes U(1)_L \otimes U(1)_\Sigma \otimes U(1)_G \times P_R$. $B$ and $L$ are the usual baryon and lepton numbers, under which only the matter fields of the MSSM are charged, the new fields charged under $SU(2)_s$ are taken to be neutral.  $G$ is a conserved quantum number for the G-quarks, $P_R$ is the usual R-parity. With this superpotential, $\Phi, \bar \Phi, \Sigma_u$ and $\Sigma_d$ can be all even or all odd under $P_R$, and similarly the parity of $D_g$ is unassigned.  However, to avoid spontaneous breaking of R-parity, the components of $H$, $\Sigma$, and $\Phi$ scalars which acquire a vev must be even under R-parity.  Finally, $U(1)_\Sigma$ is a global symmetry under which $\bar \Phi, \Sigma_d$ have charges $-1/2$, while $\Phi, \Sigma_u$ have charges $+1/2$.  

When $SU(2)_s$ is broken by $\Phi$ and/or $\bar \Phi$ vevs, the $SU(2)_s \otimes U(1)_\Sigma$ is broken to a global $U(1)_s$, which we refer to as a sister charge. Under this residual symmetry, one component of $\Phi$ is charged, while the component getting a vev is neutral, and similarly for $\bar \Phi$. Likewise, one $SU(2)_L$ doublet of $\Sigma_d$ mixes with $H_d$ and is neutral under $U(1)_s$, while another is charged and does not mix, and again with $\Sigma_u$ and $H_u$.   Thus, in this theory, with no additional operators, the lightest baryon, lepton, R-parity odd field, sister-charged field and $G$-quark are all stable.   The charges of the fields are summarized in Table~\ref{tab:globalsymms}.

\begin{table}[t]
\centering
\begin{tabular}{c|c|c|c|c} 
 & $U(1)_B$ & $U(1)_L$ & $U(1)_\Sigma$ & $U(1)_G$ \\ 
 \hline  
 $Q,U^c,D^c$ & $\pm1/3$ & $0$ & $0$  & $0$ \\
 $L,E^c$ & $0$ & $\pm 1$ & $0$ & $0$  \\
$\Sigma_u$     &  $0$ & $0$ & $1/2$ & $0$  \\
$D_g$   &  $0$ & $0$ & $0$ & $1$  \\
$\Sigma_d$      &  $0$ & $0$ & $-1/2$ & $0$  \\
$D_g^c$     &  $0$ & $0$ & $0$ & $-1$  \\
$\Phi$             &  $0$ & $0$ & $1/2$ & $0$  \\
$\bar{\Phi}$      &  $0$ & $0$ & $-1/2$ & $0$  \\
\end{tabular}
\caption{Charges of fields under the global symmetries.  The non-MSSM fields are also charged under a gauged $G_s$.}
\label{tab:globalsymms}
\end{table}

The presence of additional operators can break the symmetries further.  In particular it is possible to break R-parity in several ways.  Below we discuss a particularly interesting case, with novel phenomenology, that breaks R-parity in a new way in the lepton sector, but before doing so we discuss an alternative approach, with less novel phenomenology, of R-parity breaking in the hadron sector.

\subsection{Hadronic RPV}\label{sec:hadRPV}

It is simple to break $U(1)_G$, for instance by including a small mass-mixing term\footnote{We include this rather than $\Phi D_g D^c$ at least initially, as $D_g^c$ is considered to be in a multiplet with $\Sigma_d$ and thus to have the same $G_s$ representation. Assuming the presence of even an approximate $U(1)_{\Sigma}$, $\bar{\Phi} D_g D^c$ should be more significant than ${\Phi} D_g D^c$.}
\be
\bar{\Phi} D_g D^c.
\label{eq:gquarkmix}
\ee This operator has charge under $B$, $G$ and $\Sigma$, and its inclusion breaks one linear combination of these symmetries.  The two unbroken symmetries can be taken to be the combinations $B'=B+G/3$ and $\Sigma'=\Sigma+G/2$.  \footnote{Note that, alternatively, we could have included the operator $Q D_g^c \Sigma_d$, which would have the same effect, although it breaks the three charges down to a different linear combination.} Under this modified baryon symmetry, $B'$, the field $D_g$ carries charge $1/3$.  This will allow $G$-quarks to decay
, assuming that there is some lighter particle with the residual $U(1)_s$ charge (such as $\Sigma$ or $\Phi$). 

Another operator which affects the $U(1)_G$ is $D_g^c D_g^c U^c$. This operator is charged under $U(1)_G$ and $U(1)_B$, and thus breaks a combination.  We can think of the remaining symmetry as a modified baryon number, $B''=B-G/6$, with this assignment $D_g$ has new baryon number $-1/6$. Note that the $G$-quarks {\em cannot} decay with just this operator, since there are no states lighter than them with $B''=\pm1/6$. Equivalently, it retains a $G$-quark parity. Including this operator together with (\ref{eq:gquarkmix}) breaks $U(1)_G \otimes U(1)_\Sigma \otimes U(1)_B$ down to a single $U(1)$. We can think of the residual number as a baryon number, under which $D^c, D_g, \Phi$ have charges of $-1/3, -1/6$ and $-1/2$. 

When $\bar{\Phi}$ acquires a vev it breaks $SU(2)_s \otimes U(1)_\Sigma$ down to $U(1)_s$, which we can now identify as baryon number, $B^{\prime\prime\prime}=B''-(T_s^3+\Sigma^\prime)$.  That is, different components of $D_g$ under the $SU(2)_s$ transformation have different baryon numbers. Simply put, since the operator is $D_g^{c1} D_g^{c2} U^c$ (with 1,2 indexing the $SU(2)_s$ component), if we set $D_g^{c1}$ as the field that mass mixes, then we can treat $D_g^{c2}$ as a diquark, and the residual sister symmetry as baryon number. Since, in a similar fashion to the lepton RPV of Section~\ref{sec:rpv}, we can redefine baryon number and therefore R-parity to include the sister charge, this operator does {\em not} allow the LSP to decay as a consequence.

We can achieve R-parity violation in this fashion if we include the operator $\Phi D_g D^c$, along with $\bar{\Phi} D_g D^c+D_g^c D_g^c U^c$, which explicitly breaks $U(1)_\Sigma$. In essence, this mass mixes both components of the $G$-quarks, so that $D_g^c D_g^c U^c$, in the mass eigenstate basis, yields the usual $UDD$ RPV operator. Alternatively, if there is an additional ${\bf 5 + \bar 5}$ uncharged under $SU(2)_s$, we could integrate it out and leave the operator 
\be
\frac{1}{M_{5 \bar 5}} \bar \Phi D_g^c D^c U^c.
\ee
In the presence of (\ref{eq:gquarkmix}), this acts as the usual baryon number violating operator $UDD$. Interestingly, this preserves $U(1)_s$ and thus the lightest sister particle (or LSiP) remains a potential dark matter candidate.

Finally, we comment that including {\em both} terms such as $\Sigma_d \Sigma_d E^c$, discussed in section~\ref{sec:lepRPV}, as well as $\bar \Phi D_g D^c + D_g^c D_g^c U^c$ can lead to $B+L$ violating (but $B-L$ conserving and $R$-parity conserving) operators such as $U D^c D^c_g E H_d \Sigma_d$. Since these operators preserve $B-L$, they will not lead to proton decay, but could lead to neutron decay inside a nucleus. While the constraints on such operators are weaker than with $B-L$ violation, they are clearly constrained, and we leaved detailed studies for this hybrid scenario for future study.

\subsection{Leptonic RPV}\label{sec:lepRPV}

Within the MSSM, the operator $H_d H_d E^c$ is never discussed, and for good reason - since the $SU(2)_W$ indices of $H_d$ are contracted antisymmetrically, it vanishes identically. However, as soon as there is an additional flavor of $H_d$, the operator $H_d^1 H_d^2 E^c$ can be present. This sister-Higgs R-parity violation is qualitatively different from usual RPV in the MSSM, as we shall discuss. Let us first see how this arises in the more complicated scenario where the sister Higgs is an $SU(2)_s$ doublet.

We can naturally include the superpotential operator 
\be
\kappa_i \Sigma_d \Sigma_d E_i^c~.
\label{eq:shrpv}
\ee 
This breaks $U(1)_\Sigma\otimes U(1)_L$ down to $U(1)_{L'}$ where $L'=L-\Sigma$ and under which $\Sigma_d$ carries lepton number $1/2$.  In addition this operator determines a unique R-parity assignment within the $\Sigma$ fields that leaves R-parity preserved.  When $\Phi$ acquires a vev, we can think of the residual sister charge as identified with lepton number.  Hence, this operator does not break R-parity without some additional effect.  More explicitly, there is a new, preserved, R-parity we can define, $R'=(-1)^{3(B-Q_s)+2\mathbf{s}}$, where $Q_s=T_s^3+L'$ and $\mathbf{s}$ is the spin of the particle.  The simplest modification that leads to R-parity violation is to include the terms $\bar \Phi H_u \Sigma_d$ and/or $\Phi H_d \Sigma_u$ in the superpotential. These explicitly break $U(1)_\Sigma$, which means there is no residual $U(1)_s$, and thus both components of $\Sigma_d$ will ultimately mix with Higgs bosons.  There are no stable states in the theory that may act as dark matter.

Alternatively, if again we consider a ${\bf 5 + \bar 5}$ uncharged under $SU(2)_s$, we could integrate it out and leave the operator 
\be
\frac{\kappa'_i}{M_{5 \bar 5}} \Phi \Sigma_d H_d E_i^c~.
\label{eq:rpvhigherdim}
\ee
This operator preserves $U(1)_\Sigma$, and hence $U(1)_s$, but it is not possible to define a preserved R-parity in the presence of (\ref{eq:SHsup}) and (\ref{eq:rpvhigherdim}).  This operator breaks R-parity (and lepton number), and thus we are left with an interesting possibility: that R-parity is broken, but that a residual global symmetry is intact, leaving a potential candidate for dark matter.  In this case the collider production of DM will either be through electroweak processes, or through the decay of the G-quarks which have QCD production, but may be very heavy and so kinematically suppressed.

In either of these two cases, stable DM candidate or not, there are interesting implications for collider phenomenology.  R-parity violation (RPV) will remove the usual $\met$ signature of SUSY, but, unlike hadronic RPV, where the $\met$ is converted to jets (and can thus be extremely challenging to discover), here, the $\met$ is converted to leptons+jets, which is much more tractable. Furthermore, under the assumption that some flavor structure dictates the sizes of couplings, the final state lepton is essentially all $\tau$.  Then the final state will contain non-negligible $\met$ and the SUSY signatures, while more challenging, are not necessarily impossible.  If the dominant decays are to $e,\,\mu$ then the $\met$ requirement of many new physics searches may not be met and the constraints will be weaker.  We discuss below the possible signatures for decays for various LSPs in the presence of this form of leptonic RPV, and outline which existing searches may be sensitive to these signatures.  We leave a detailed study of efficiencies and resulting bounds for future work~\cite{ustodo}.

\subsection{Constraints on Sister RPV}

Both (\ref{eq:shrpv}) and (\ref{eq:rpvhigherdim}) will lead to mixing between the RH leptons and the Higgsinos. We concentrate for concreteness on the effects of (\ref{eq:shrpv}) with one or both of $\Phi H_d \Sigma_u +\bar \Phi H_u \Sigma_d$.  One linear combination of leptons, the combination $\sum \kappa_i \ell_i$, will have a mass contribution $\sum \kappa_i^2 v^2 \cos^2\beta/M_\Sigma$.  
Unlike the conventional $L H_u$ and $LLE^c$ RPV operators these contributions do not affect the left-handed leptons, meaning that constraints from neutrino mass are much weaker. Likewise, since there is only a single lepton appearing in this operator, constraints on dilepton resonances are not expected to be relevant. \footnote{The exception is if the LSP is a sneutrino, which can decay to two leptons.}

Rare decays are expected, but because this operator couples to right-handed leptons only, the transition will have a chiral suppression, i.e., we expect a dipole operator for $\ell_i \rightarrow \ell_f \gamma$ ($\ell_i = \mu, \tau$, $\ell_f = \mu,  e$) of $D_{if}\sim m_\ell \kappa_i \kappa_f e/16 \pi^2 M_\Sigma^2$.  Current bounds on the lepton flavor violating branching ratios place constraints on the $\kappa_i$ of,
\be
|\kappa_\mu \kappa_e| \ltap 3\times 10^{-6} \left(\frac{M_\Sigma}{100\,\gev}\right)^2 \quad \text{and}\quad
|\kappa_\tau \kappa_\ell| \ltap 2\times 10^{-4} \left(\frac{M_\Sigma}{100\,\gev}\right)^2~.
\ee


\subsection*{Squark LSP}

\begin{figure}[t] 
   \centering
   \includegraphics[width=0.4\textwidth]{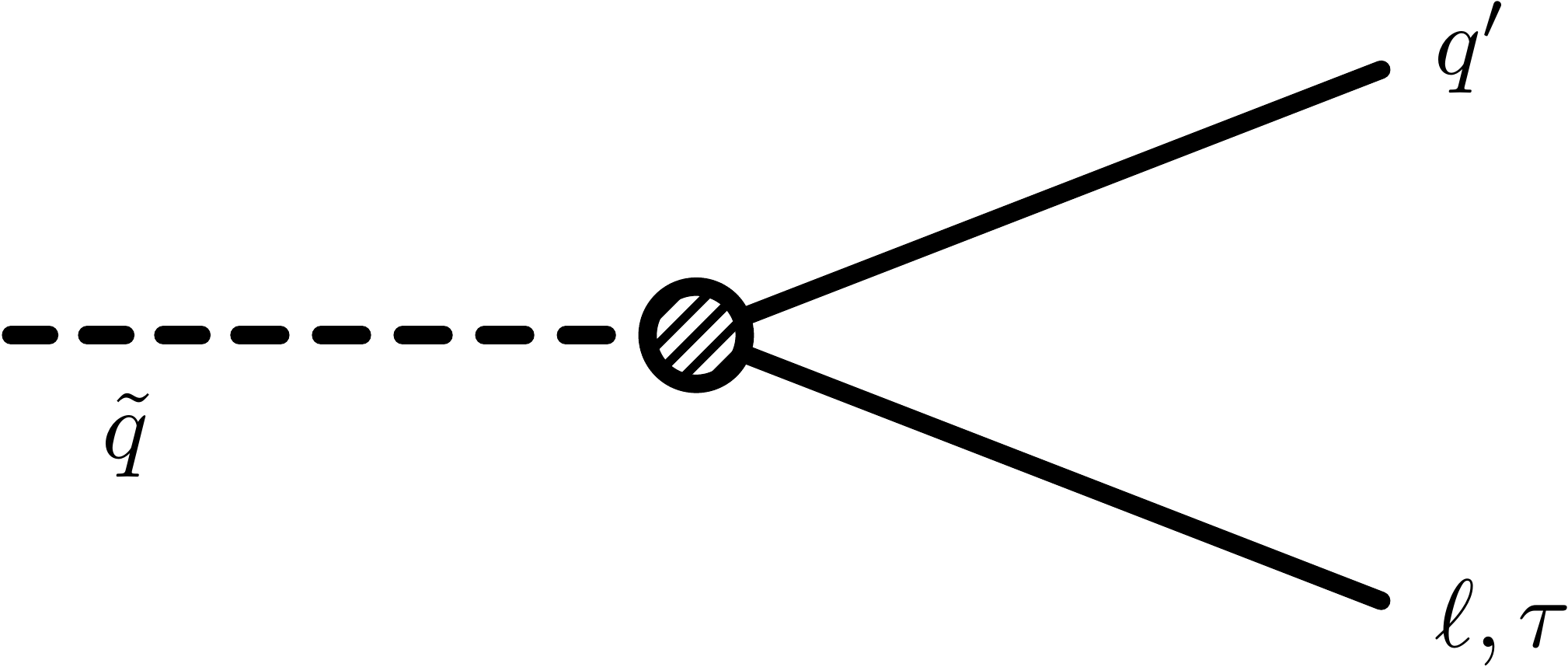} 
    \includegraphics[width=0.4\textwidth]{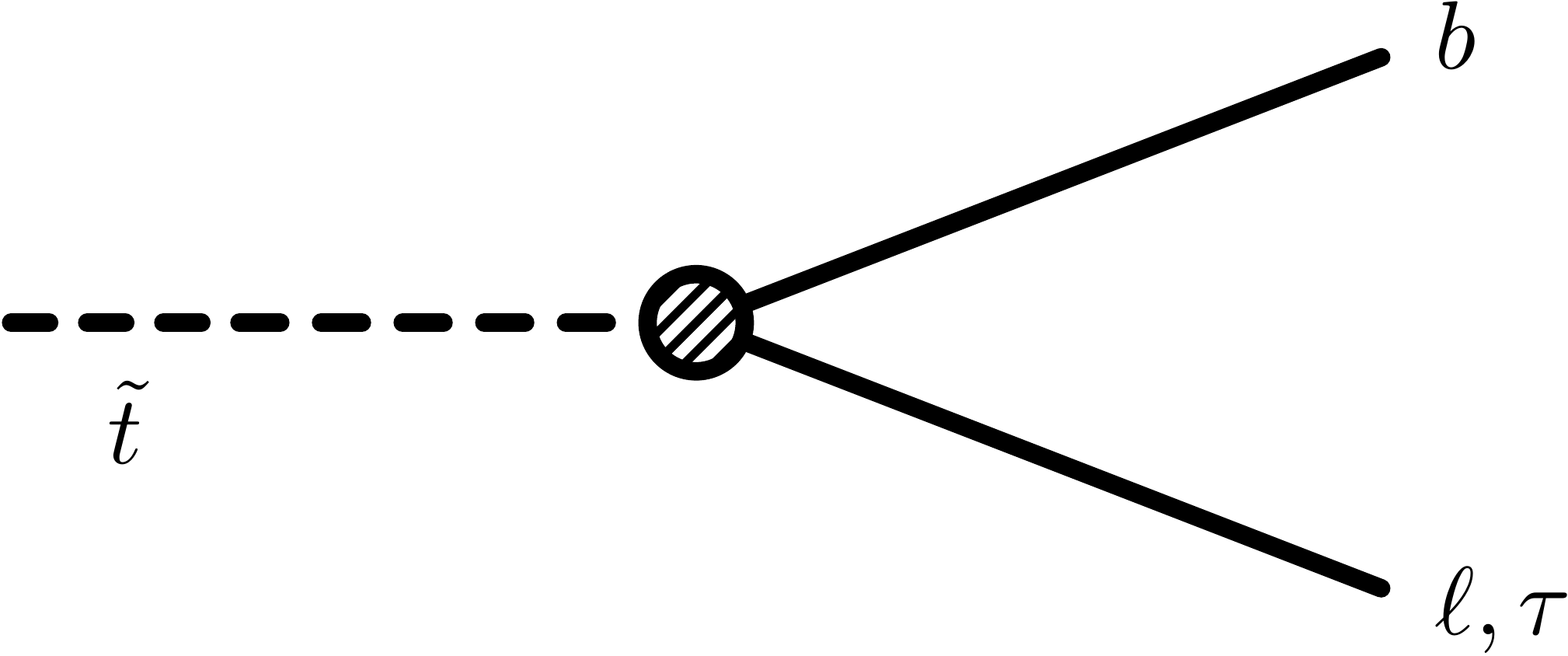}
   \caption{Squark and stop decay modes in the leptonic sister RPV scenario.}
   \label{fig:squarkdecay}
\end{figure}

A squark LSP in the presence of sister RPV is an interesting case to consider. Given that the operator $H_d^1 H_d^2 E_i^c$ mixes the chargino with a charged lepton, the dominant decay mode of a squark LSP may be $\tilde{q}\rightarrow q^\prime \ell^\pm$ ( Fig.~\ref{fig:squarkdecay}), similar to the decay induced by the RPV operator $Q L D^c$ in the MSSM. Note, however, that contrary to $Q L D^c$, the sister RPV operator $H_d^1 H_d^2 E_i^c$ does not induce the decay $\tilde{q}\rightarrow q \nu_\ell$ at tree-level.

If the squarks belong to the first or second generation and dominantly decay to $\tilde{q}\rightarrow q^\prime e^\pm$ or $\tilde{q}\rightarrow q^\prime \mu^\pm$, straightforward bounds on $m_{\tilde{q}}$ can be directly extracted from leptoquark searches. ATLAS carried out searches for first and second generation leptoquarks with $1~\text{fb}^{-1}$ of luminosity \cite{Aad:2011ch,ATLAS:2012aq}, and constrained the masses of leptoquarks to be $m_{LQ}>685~\GeV$ for $LQ\rightarrow \mu j$ and $m_{LQ}>660~\GeV$ for $LQ\rightarrow e j$. CMS also searched for second generation leptoquarks with $2~\text{fb}^{-1}$ of luminosity \cite{CMS-PAS-EXO-11-028}, and placed the bound $m_{LQ}>632~\GeV$ for $LQ\rightarrow \mu j$.

The bounds above on leptoquark masses directly apply to squarks decaying to those same final states, since the production cross section for squarks and leptoquarks are effectively the same in the limit in which the gluino is heavy and has a negligible contribution to squark pair production. However, in generic scenarios in which the first and second generation of right and left handed squarks are degenerate, the squark pair production cross section is a factor of 8 larger than leptoquark pair production. It is straightforward to extract the limit on squark masses in this case: the ATLAS search bounds the production cross section of $\tilde{q}\rightarrow q^\prime e^\pm$ to be $\sigma(pp\rightarrow \tilde{q}\tilde{q}^*\rightarrow e^+ e^- jj)\leq 5\times 10^{-3}~\text{pb}$, implying that $m_{\tilde{q}}\gsim 880~\GeV$. Similarly, the strongest constraint on $\tilde{q}\rightarrow q^\prime \mu^\pm$ comes from CMS. It bounds the production cross section to be $\sigma(pp\rightarrow \tilde{q}\tilde{q}^*\rightarrow \mu^+ \mu^- jj)\leq 3\times 10^{-3}~\text{pb}$, which translates into $m_{\tilde{q}}\gsim 900~\GeV$.

Those searches cannot, however, be as straightforwardly applied to the case in which the dominant squark decay mode is  $\tilde{q}\rightarrow q^\prime \tau^\pm$. In that case searches specifically targeting new physics signatures with jets and two tau leptons place stronger bounds. ATLAS performed a search for events with jets, at least two tau leptons and missing energy using $2~\text{fb}^{-1}$ of data \cite{ATLAS:2012ag}. We re-interpreted the results of those searches in order to place bounds on first and second generation squarks decaying as $\tilde{q}\rightarrow q^\prime \tau^\pm$, as well as stop LSPs decaying as $\tilde{t}\rightarrow b \tau^+$. We used \texttt{MadGraph5} to simulate $pp\rightarrow \tilde{q}\tilde{q}^*$ and $pp\rightarrow \tilde{t}\tilde{t}^*$ at the 7 TeV LHC. Those events were passed through \texttt{Pythia6} for decay, showering and hadronization, and then handed to \texttt{PGS4} for a rough simulation of the detector response.

\begin{figure}[t] 
\includegraphics[width=7in]{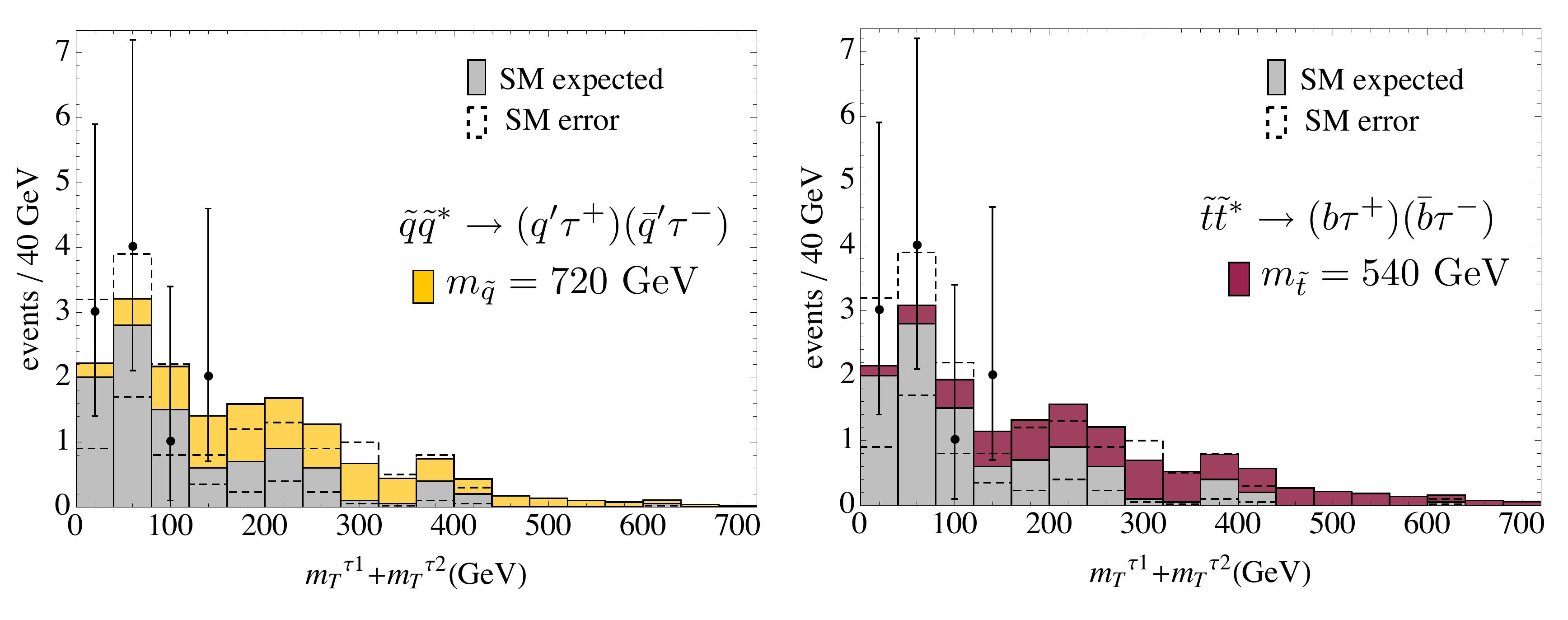}
\caption{Distribution of $m_T^{\tau_1}+m_T^{\tau_2}$ for the event selection of the ATLAS search in  \cite{ATLAS:2012ag}.  The SM background, as estimated in  \cite{ATLAS:2012ag}, is displayed in gray, and the measured data is shown as black dots. Two benchmark signals with masses at the boundary of the this analysis' reach are displayed as well: (left) degenerate 1st and 2nd generation squark LSP's decaying as $\tilde{q}\rightarrow q^\prime \tau^\pm$, and (right) stop LSP decaying as $\tilde{t}\rightarrow b \tau^+$.}
\label{fig:squarkLSP}
\end{figure}

The ATLAS search in \cite{ATLAS:2012ag} selected events with at least two jets, the leading jet with transverse momentum $p_T>130~\GeV$ and the sub-leading jet with transverse momentum $p_T>30~\GeV$. In addition, it required two hadronic tau candidates with $p_T>20~\GeV$ and missing transverse energy $\met>130~\GeV$. It also placed a cut $\Delta\phi(\vec\met,\vec p_{T1,2})>0.4$ between the missing transverse momentum and the two leading jets, and $m_{\text{eff}}>700~\GeV$. In Figure \ref{fig:squarkLSP} we display the distribution of events in terms of the sum of the transverse mass of the two tau candidates, for a few benchmark signals. The signal region chosen by ATLAS corresponds to $m_T^{\tau_1}+m_T^{\tau_2}>80~\GeV$, and it bounds the number of signal events in that region to be $\lsim6$ events. For the scenario of degenerate first and second generation squarks decaying as $\tilde{q}\rightarrow q^\prime \tau^\pm$, that excludes squark masses $m_{\tilde{q}}\lsim720~\GeV$. For the stop scenario with decay $\tilde{t}\rightarrow b \tau^+$ the exclusion is $m_{\tilde{t}}\lsim540~\GeV$. This bound is surprisingly similar to the one obtained by the CMS dedicated search to this signature \cite{CMS-EXO-12-002}, which constrains stop masses to be $m_{\tilde{t}}\lsim525~\GeV$. Searches for charged higges from stop decays, $t\rightarrow b H^+ \rightarrow b(\tau^+\nu)$, such as \cite{Aad:2012tj}, also yield non-trivial bounds. We estimate that the bound from this particular search to be $m_{\tilde t}\lsim 300~\GeV$, hence weaker than the bound from \cite{ATLAS:2012ag, CMS-EXO-12-002}.

\subsection*{Neutralino LSP}\label{sec:neutralinoLSP}

In the presence of the leptonic sister RPV operator $H_d^1 H_d^2 E_i^c$, a neutralino LSP can decay to $\ell^\pm H^\mp$. As we discussed before, the phenomenology of the light Higgs sector is that of a type-I 2HDM, and therefore $H^\pm$ will decay dominantly to $t\bar b$ (or $Wb\bar b$), $\tau\nu$ and $c\bar s$, with branching ratios determined by its mass and whether it is allowed to be on-shell or not. If $H^\pm$ is too heavy, the decay $\chi_1^0\rightarrow\ell^\pm W^\mp$ might dominate. Either way, the expected signatures from neutralinos at the bottom of SUSY cascades should contain a large number of leptons. If these neutralinos are being produced through strong interactions (such as gluinos or squarks), these scenarios should be significantly constrained by same-sign dileptons \cite{CMS:2012th,Chatrchyan:2012sa,ATLAS:2012ai,CMS-PAS-SUS-12-017,ATLAS-CONF-2012-069} and multilepton searches \cite{Chatrchyan:2012ye,ATLAS-CONF-2012-077,ATLAS-CONF-2012-035,ATLAS-CONF-2012-001,ATLAS-CONF-2011-158,ATLAS-CONF-2011-144}, which have very low backgrounds and hence have very powerful sensitivity even for models with very low branching ratio to multileptons. Other potentially constraining searches are same-sign tops (if the decay $\chi_1^0\rightarrow\ell^\mp H^\pm\rightarrow\ell^\mp (t b)^\pm$ is dominant), first and second generation leptoquarks  \cite{Aad:2011ch,ATLAS:2012aq,CMS-PAS-EXO-11-028}, or GMSB with LFV \cite{Collaboration:2012yw}.

\begin{figure}[t] 
   \centering
   \includegraphics[width=0.4\textwidth]{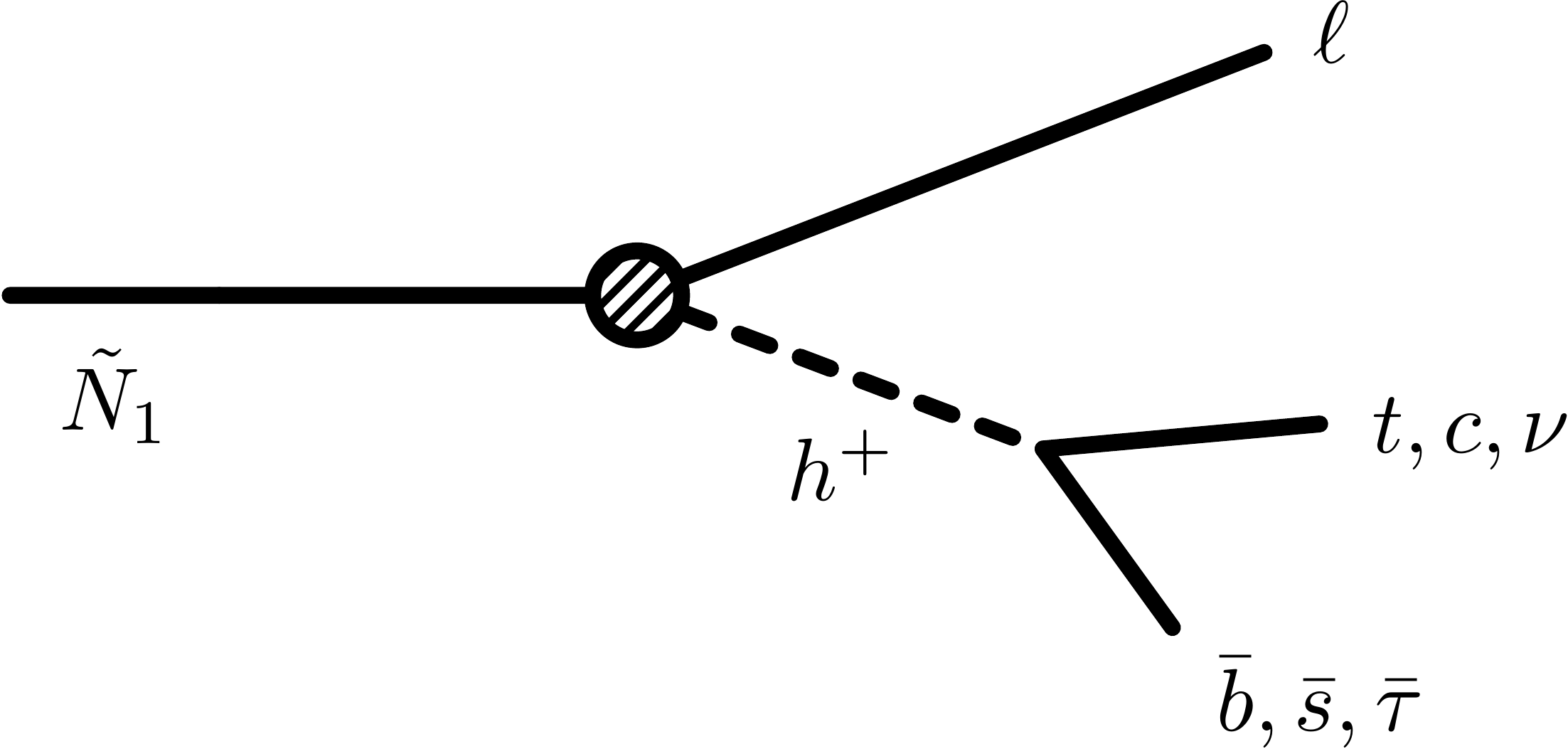} 
    \includegraphics[width=0.4\textwidth]{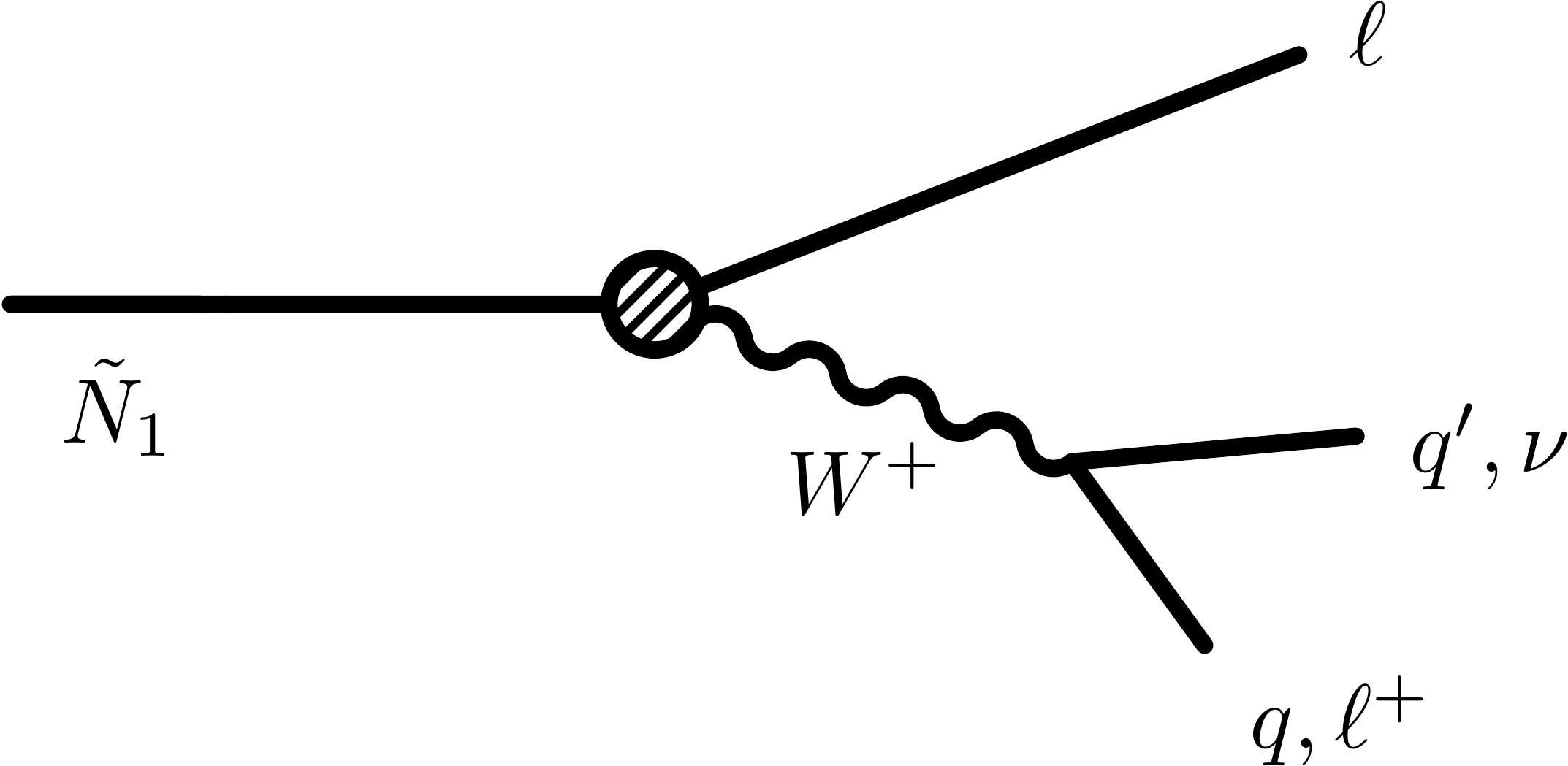}
   \caption{Neutralino LSP decay modes induced by leptonic sister RPV.}
   \label{fig:neutdecay}
\end{figure}

\subsection*{Chargino LSP}

\begin{figure}[t] 
   \centering
   \includegraphics[width=0.3\textwidth]{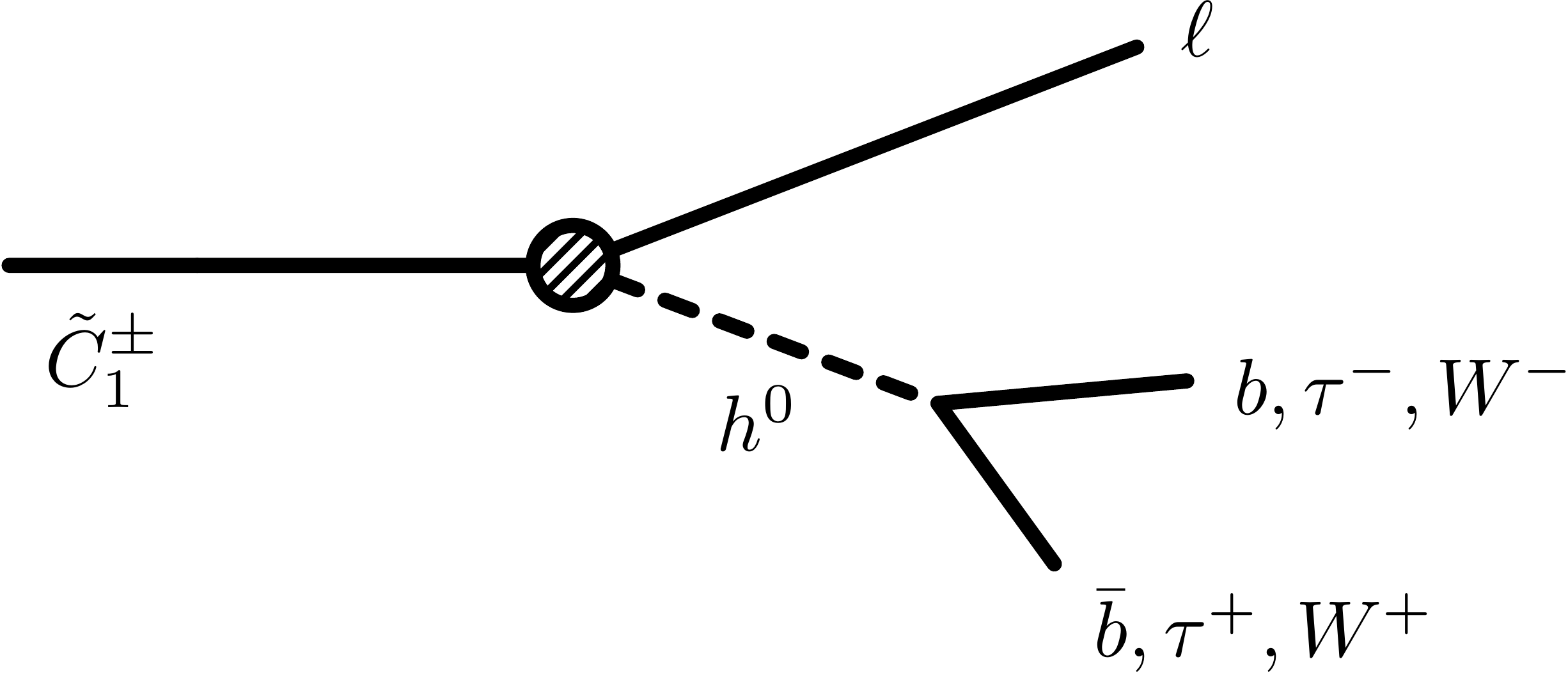} 
    \includegraphics[width=0.3\textwidth]{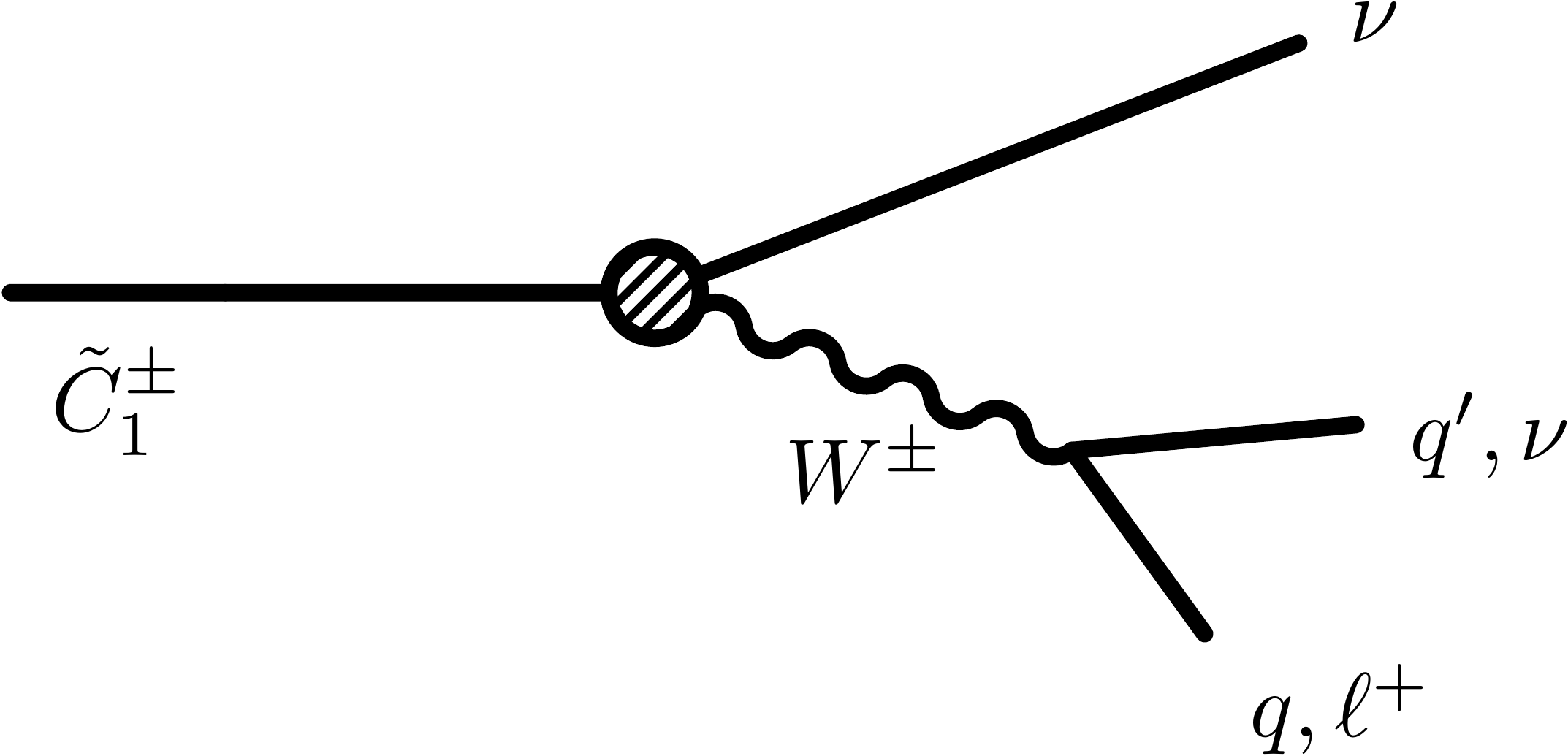}
    \includegraphics[width=0.3\textwidth]{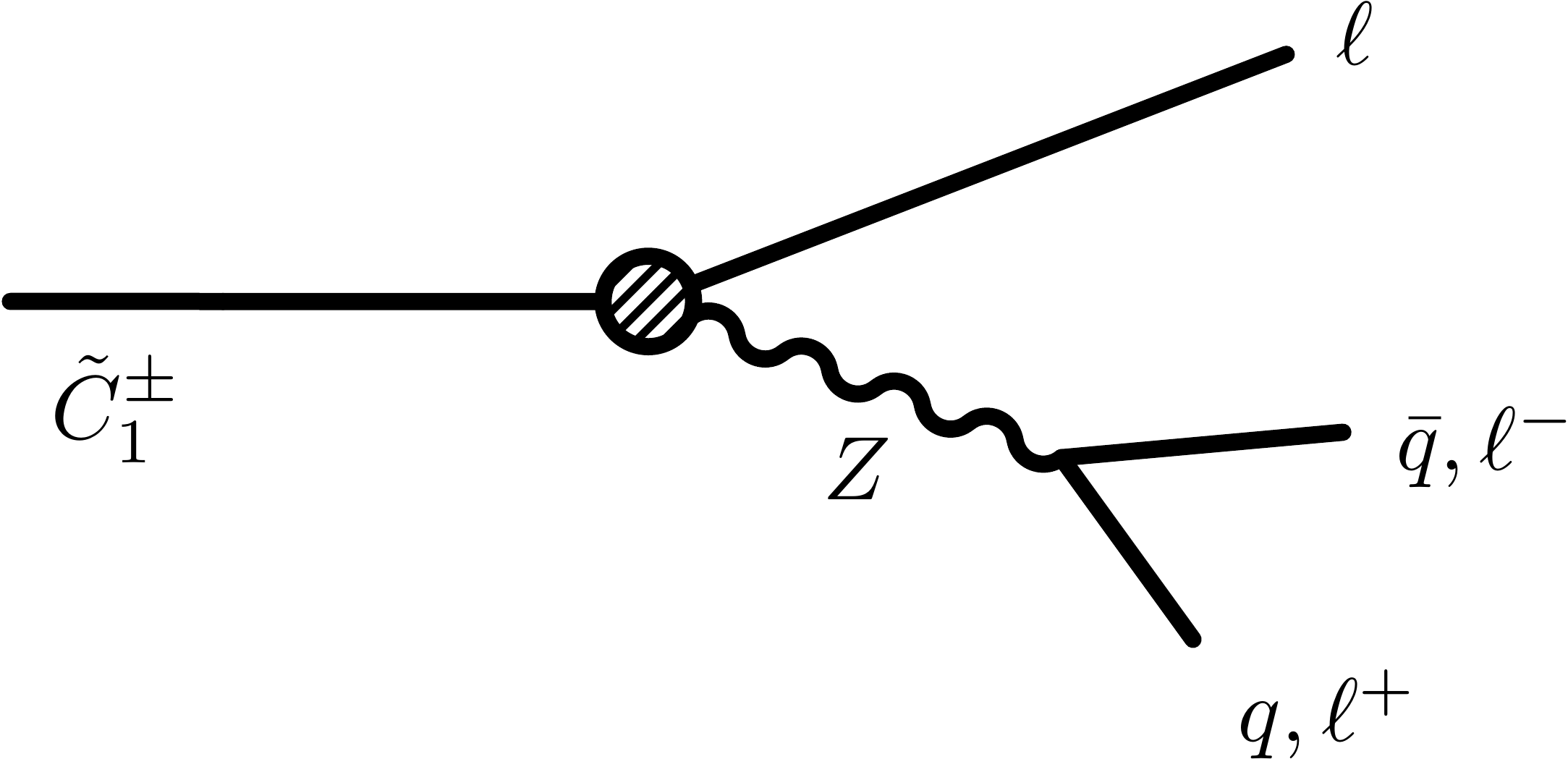}
   \caption{Chargino LSP decay modes induced by leptonic sister RPV.}
   \label{fig:chargdecay}
\end{figure}

It is also interesting to consider the situation where the chargino is the LSP (see \cite{Kribs:2008hq} for a discussion of scenarios), or when its decay to a neutralino is suppressed due to phase space. In this case, the decay $\chi^\pm \rightarrow\ell^\pm h_x^0 $ will be quite challenging, see Figure~\ref{fig:chargdecay}. If $h_x$ couples to up quarks, it is possible that we would get a sizeable signal of $WW$ (through its mixing with the Higgs), but if unavailable, then $c \bar c$ would dominate. Or, if coupling through down quarks, $b \bar b$, which can perhaps aid by adding b-tags along with hard leptons (possibly $\tau$s).  Alternative decays to $W^\pm\nu$ or $\ell^\pm Z^0$ should be constrained constrained from standard SUSY searches in jets +$\met$ and/or multileptons.

\subsection*{Slepton/Sneutrino LSP}

\begin{figure}[h] 
    \includegraphics[width=0.8\textwidth]{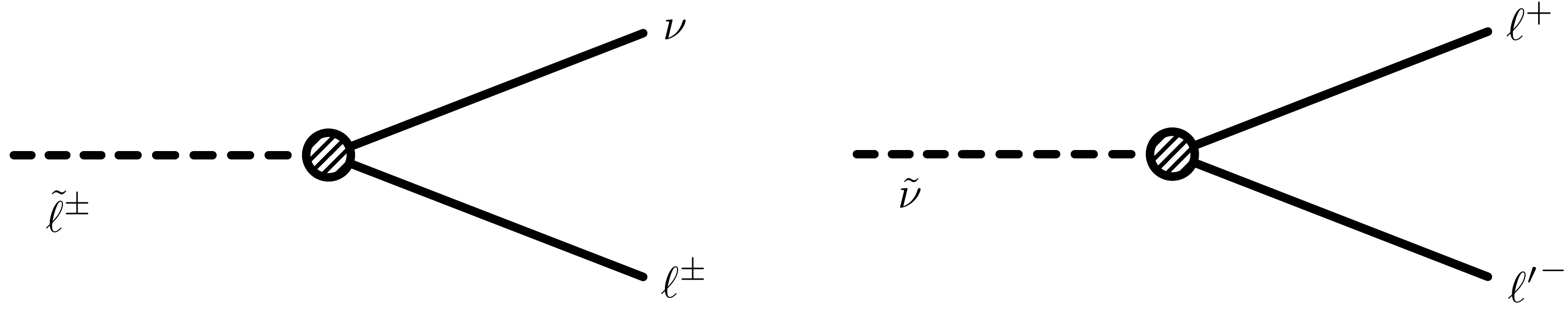}
   \caption{Slepton and Sneutrino LSP decays induced by leptonic sister RPV}
   \label{slepsnu}
\end{figure}

Slepton or sneutrino LSPs decay in a similar fashion as in standard leptonic RPV induced by the $LLE$ operator, see Fig.\ref{slepsnu}.  Searches for direct slepton production have been recently released by ATLAS \cite{ATLAS-CONF-2012-076} and constrain sleptons decaying as $\tilde\ell^\pm\rightarrow \ell^\pm \nu$ (where $\ell^\pm=e^\pm$ or $\mu^\pm$) to be $m_{\tilde\ell^\pm}\gsim185~\GeV$. These searches, however, cannot constrain sneutrino pair-production due to the absence of missing energy. However, multi-lepton searches requiring 4 leptons should have sensitivity to $\tilde\nu\tilde\nu \rightarrow 4\ell$.

\subsection*{Gluino LSP}

Gluino LSPs will decay to dijets plus a charged lepton. If the charged lepton is a $\tau^\pm$, standard jets+$\met$ plus 0, 1 or 2 charged leptons should have sensitivity to pair-produced gluinos. For gluinos decaying to jets plus  $e^\pm$ or $\mu^\pm$, however, the absence of missing energy makes this signature more challenging. Limits from current searches on leptoquarks, black-holes or same-sign dileptons may be re-cast for $\tilde g \tilde g\rightarrow (qq\ell^\pm)(qq\ell^\pm)$; however dedicated searches may have a potentially much higher reach.

\section{Dark Matter}
\label{se_c:dm}

One appealing element of SUSY is the presence of a dark matter candidate in the form of the LSP. Under the assumption of R-parity conservation, the neutralino often serves as a WIMP candidate. In this theory, because of the sister Higgsinos, the state forming the WIMP is more complicated, with the new forces altering the DM freezeout story. While a thorough study of dark matter is beyond our scope, some discussion is clearly worthwhile.

\subsection{Dark Matter with R-Parity Violation}
In general, there is a tradeoff in models between the presence of RPV (which can be an explanation of the so-far elusive nature of supersymmetry) and a dark matter candidate. The standard lore is that since the LSP is the dark matter candidate, violating R-parity (and thus removing the $\met$ signal) will remove a cosmologically stable particle from the theory. However, within the sister MSSM, that lore is easily violated by the presence of a residual global symmetry.

In the Sister Higgs scenario, that symmetry can be the global $U(1)_s$, which, if preserved, makes the lightest state charged under this symmetry (the ``LSiP") stable and hence a potential dark matter candidate. If the $U(1)_s$ is broken down to a $Z_2$, then the LSiP is Majorana. As we shall see, the Dirac case is quite constrained by direct detection, but the Majorana case is much less so. 

Among the fermions carrying $U(1)_s$ charge, there are three Dirac states comprised of $\tilde{\phi}_s \equiv ( \tilde{\Phi}_1,\tilde{\bar{\Phi}}_2)$, 
$\tilde{\Sigma}_s \equiv (\tilde{\Sigma}_{u1},\tilde{\Sigma}_{d2})$, and $\tilde{W}_s \equiv (\tilde{W}_{s^+},\tilde{W}_{s^-})$.
The masses and mixings of those fields can be extracted from:
\begin{equation}
\label{sistermassmatrix}
\begin{array}{ccc}
\mathcal{L}~\supset~[\tilde{\Phi}_1 & \tilde{\Sigma}_{u1} & \tilde{W}_{s^+}] \\
&&\\
&&
 \end{array}
\left[\begin{array}{ccc}
-\mu_\phi & -\lambda_u v s_\beta & g_s \langle\Phi_2\rangle \\
0 & \mu_\Sigma & 0 \\
g_s \langle\bar{\Phi}_1 \rangle& g_s v c_\beta & -M_{s}
\end{array}\right]
\left[\begin{array}{c}\tilde{\bar{\Phi}}_2 \\ \tilde{\Sigma}_{d2} \\ \tilde{W}_{s^-} \end{array}\right].
\end{equation}

Those fields couple vectorially to the Standard Model $Z^0$ boson, either directly though their $\Sigma_s$-components, or through $Z^0-Z^0_s$ mixing. If the lightest mass eigenstate is the dark matter, its coupling to the $Z^0$ is constrained by direct detection experiments. The LSiP-{\it nucleon} cross section induced by $Z^0$ t-channel exchange is:
\begin{equation}
\sigma_n\simeq\frac{1}{4\pi}\left(\kappa_{\text{LSiP}-Z}\frac{g_2}{c_w}\frac{1}{m_Z^2}\right)^2\mu_n^2\frac{(A-Z)^2}{A^2},
\end{equation}
where $\kappa_{\text{LSiP}-Z}$ is the dark matter coupling to $Z^0$, $\mu_n$ is the dark matter-nucleon reduced mass and A (Z) is the nucleus atomic mass (number). This cross section is constrained by XENON100 \cite{Aprile:2011hi} to be $\sigma_n\lsim10^{-44}\text{cm}^2$. This translates into a bound on the dark matter-$Z^0$ coupling of $\kappa_{\text{LSiP}-Z}\lsim3.7\times10^{-4}$, which rules out $\tilde{\Sigma}_s$ as a Dirac dark matter candidate since it is charged under $SU(2)_W$ and its coupling to $Z^0$ ($\kappa_{\tilde{\Sigma}_s-Z^0}={g_2}/{2c_w}$) violates the direct detection bound by three orders of magnitude.
This bound also severely constrains $\tilde{\phi}_s$ as a Dirac dark matter candidate, since it mixes significantly with $\tilde{\Sigma}_s$ through $\lambda_u \langle H_u\rangle \Sigma_d \Phi$.

These bounds can be evaded however, by adding small Majorana mass terms, for instance:
\begin{equation}
\mathcal{L}\supset \delta m_1\tilde{\Phi}_1\tilde{\Phi}_1 + \delta m_2 \tilde{\bar{\Phi}}_2 \tilde{\bar{\Phi}}_2 ,
\end{equation}
or analogous terms for $\tilde{\Sigma}_s$. This term breaks $U(1)_s$ but preserves a $Z_2$ parity that stabilizes the dark matter. It also splits the mass eigenstates by $\delta m_1+\delta m_2$. Since those states are Majorana, their Z-mediated elastic scattering is spin-dependent, and moreover suppressed by $\delta m/\mu_\phi$. The spin-independent scattering is inelastic and hence can be suppressed or even shut off for large enough mass splittings \cite{TuckerSmith:2001hy}. Note that it is non-trivial to achieve this term as superpotential couplings $\Phi \Phi + \bar \Phi \bar \Phi$ vanish identically by the antisymmetry of the $SU(2)_s$ indices. Thus, generating this term would naturally involve SUSY breaking or higher dimension operators. We shall not pursue this except to note that this is not a simple model addition.

In the absence of such $U(1)_s$-breaking majorana masses, however, the only viable Dirac dark matter is a state that is almost purely sister gaugino, $\tilde{W}_{s}$. The direct detection bound, in this case, translates into constraints on:

\begin{itemize}
{\item the mixing of $\tilde{W}_{s}$ with $\tilde{\Sigma}_s$:
\begin{equation}
\label{WSmixing}
\frac{1}{2}\frac{g_2}{c_w}~\frac{\theta^2_{\tilde{\Sigma}_{u1}\tilde{W}_{s^+}}+\theta^2_{\tilde{\Sigma}_{d2}\tilde{W}_{s^-}}}{2}<3.7\times10^{-4},
\end{equation}
}
{\item the mixing of the Standard Model $Z^0$ with the sister gauge boson $Z^0_s$:
\begin{equation}
\label{zzsmixing}
g_s\times\theta_{Z-Z_s}=\frac{g_s^2}{\sqrt{g_2^2+g_Y^2}}\frac{m_Z^2}{m_{Z_s}^2-m_Z^2}\cos^2\beta~\lsim~3.7\times10^{-4}.
\end{equation}
}
\end{itemize}

\begin{figure}[t] 
\includegraphics[width=6.5in]{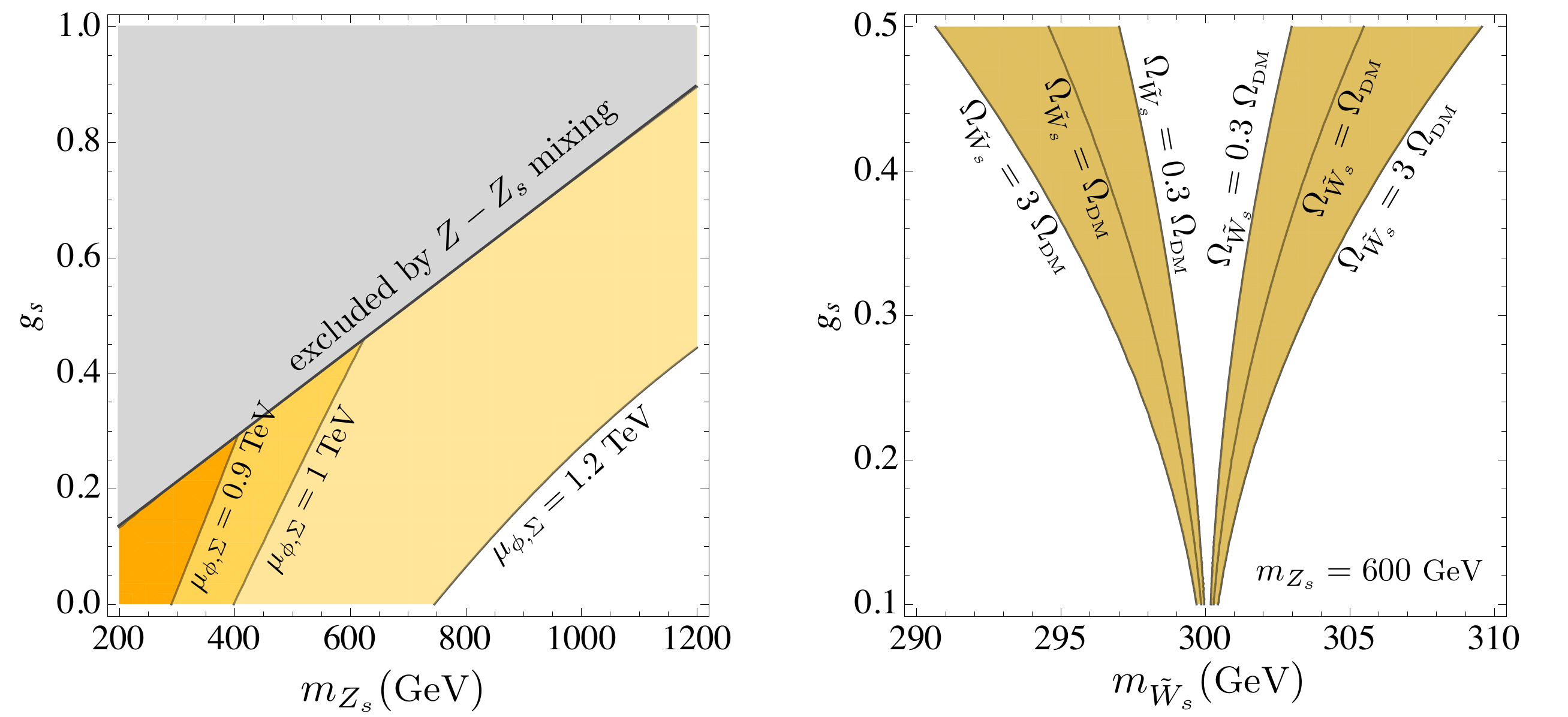}
\caption{(Left) Direct detection bounds on $g_s$ vs $m_{Z_s}$ fixing $\tan\beta$=4 and $\lambda_u$=0.6, assuming that dark matter is dominantly the gaugino LSiP, $\tilde{W}_s$. The gray region is excluded due to large $Z-Z_s$ mixing, and the colored regions are allowed when the mixing of $\tilde{W}_s$ with $\tilde{\Sigma}_s$ is sufficiently suppressed, which can be achieved with $\mu_\phi, \mu_\Sigma\sim\OO(1)$ TeV. (Right) $\tilde{W}_s$ relic abundance, assuming $m_{Z_s}=600~\GeV$.}\label{dmplot}
\end{figure}

The later constraint is illustrated in the left-hand plot of Fig.\ref{dmplot} for $\tan\beta=4$. In the gray colored region the $Z-Z_s$ mixing violates the bound in (\ref{zzsmixing}), predicting a direct detection rate larger than presently allowed. Constraint (\ref{WSmixing}) on the mixing between $\tilde{W}_{s}$ and $\tilde{\Sigma}_s$ can be satisfied with large $\mu_\phi$ and $\mu_\Sigma$ terms (see eq.(\ref{sistermassmatrix})). Fig.\ref{dmplot} illustrates the allowed regions in the $m_{Z_s} - g_s$ parameter space for $\mu_\phi=\mu_\Sigma$ = 0.9 TeV, 1 TeV and 1.2 TeV. One can see that the larger $\mu_\phi$ and $\mu_\Sigma$ are, the more $\tilde{\phi}_s,\tilde{\Sigma}_s$ are decoupled, suppressing their mixing with $\tilde{W}_s$ and allowing a larger parameter space in $m_{Z_s} - g_s$. Thus, insisting on a dark matter candidate of this type pushes us to a very unnatural corner of parameter space, since $\mu_\phi$ and $\mu_\Sigma$ are directly tied to electroweak symmetry breaking, and their TeV scale values imply large cancellations in order to get the correct $Z^0$ mass.

In order to complete our dark matter discussion, we also consider the dark matter relic abundance.
The dominant $\tilde{W}_s\tilde{W}_s$ annihilation channel will depend on whether there are states with $SU(2)_s$ interactions that are lighter than $\tilde{W}_s$, such as the scalar components of $\Phi$ or $\bar{\Phi}$, or G-quarks. If annihilation into those (on-shell) states is kinematically allowed, they will dominate due to large $g_s$-couplings. If not, then the dominant annihilation channel will be $\tilde{W}_s\tilde{W}_s\rightarrow Z_s \rightarrow Z^0 h^0$. The final state $W^\pm H^\mp$ is kinematically forbidden since the mass of the charged Higgs, controlled by $\lambda_u \mu_\phi \langle\bar{\Phi}\rangle$, is at the TeV scale. Annihilation through an s-channel $Z^0$ to Standard Model fermions is subdominant relative to the $Z^0 h^0$ final state when $Z-Z_s$ mixing is suppressed. However, for $m_{Z_s}\lsim350\ \GeV$, the annihilation cross section to SM fermions is larger than 10\% of the cross section to $Z^0 h^0$, so we include it as well in what follows.

The annihilation cross section for $\tilde{W}_s\tilde{W}_s\rightarrow Z_s \rightarrow Z^0 h^0$ is:
\begin{eqnarray}
\label{sigmavZH}
\langle\sigma v\rangle_{Z^0 h^0}&=&\frac{\pi}{6}~(\alpha_s\cos\beta\sin\alpha)^2~\xi\left(\frac{m_{Z^0}^2}{s},\frac{m_{h^0}^2}{s}\right)~\frac{s+2m_{\tilde{W}_s}^2}{(s-m_{Z_s}^2)^2+s^2\Gamma_{Z_s}^2/m_{Z_s}^2}\\
&\simeq&\frac{\pi}{6}~\alpha_s^2\cos^4\beta~\xi\left(\frac{m_{Z^0}^2}{4m_{\tilde{W}_s}^2},\frac{m_{h^0}^2}{4m_{\tilde{W}_s}^2}\right)~\frac{4m_{\tilde{W}_s}^2+2m_{\tilde{W}_s}^2}{(4m_{\tilde{W}_s}^2-m_{Z_s}^2)^2},
\end{eqnarray}
where $\alpha_s=g_s^2/(4\pi)$, $\alpha$ is the mixing angle that determines the mass eigenstates of the  CP-even higges $h^0$ and $H^0$, and we can ignore the $Z_s$-width since, in the region of parameter space we are considering, it is $<\OO(\text{MeV})$. Moreover, in the limit we are considering the CP-odd higgs is very heavy, so that $\sin\alpha\simeq-\cos\beta$. Finally,
\begin{equation}
\xi(x,y)\equiv\left(1+2(5x-y)+(x-y)^2\right)\sqrt{1+x^2+y^2-2(x+y+xy)}.
\end{equation}

Annihilation into SM fermions through s-channel $Z^0$ or $Z_s^0$ is given by
\begin{eqnarray}
\label{sigmavff}
\langle\sigma v\rangle_{f\bar{f}}&\simeq&4.9 \pi~\alpha_s^2~\cos^4\beta~\frac{m_Z^4}{(m_{Z_s}^2-m_Z^2)^2}~(s+2m_{\tilde{W}_s}^2)\left|\frac{1}{s-m_Z^2+ \frac{is\Gamma_Z}{m_Z}}-\frac{1}{s-m_{Z_s}^2+\frac{is \Gamma_{Z_s}}{m_{Z_s}}}\right|^2.
\end{eqnarray}

\begin{figure}[t] 
\includegraphics[width=4in]{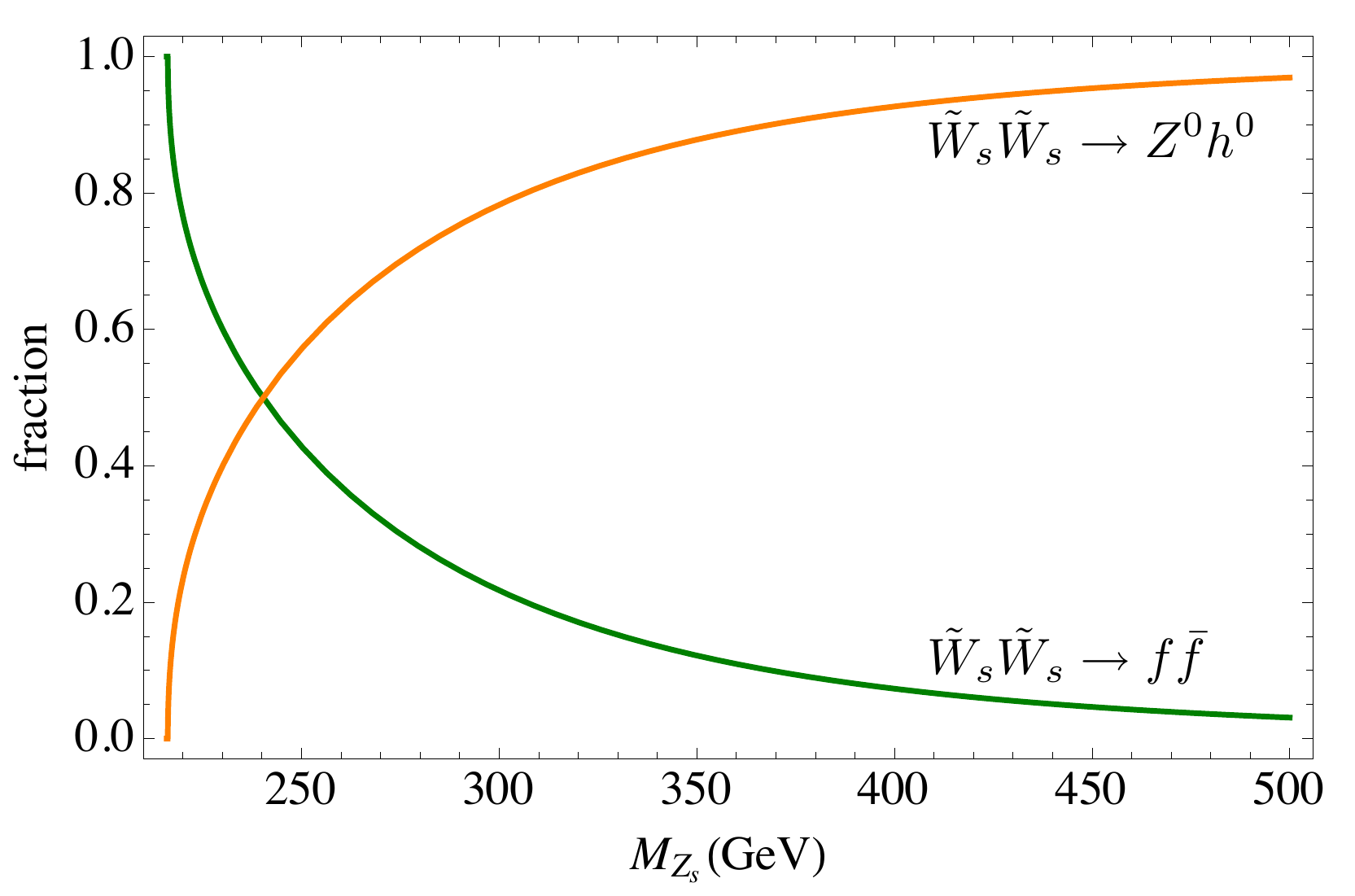}
\caption{Relative $\tilde{W}_s\tilde{W}_s$ annihilation cross section to $Z^0h^0$ vs. light SM fermions as a function of $m_{Z_s}$. The LSiP mass is set to $m_{\tilde{W}_s}=m_{Z_s}/2$.}\label{AnnihBr}
\end{figure}

From (\ref{sigmavZH}) and (\ref{sigmavff}), we see that the dominant annihilation channel ($Z^0h^0$ vs $f\bar{f}$) is determined primarily by the mass of $Z_s$ (there is a smaller dependence on $m_{\tilde{W}_s}$ that is only important near threshold). Fig.\ref{AnnihBr} displays the relative fraction between the two annihilation channels (assuming $m_{\tilde{W}_s}\approx m_{Z_s}/2$). As mentioned previously, for $m_{Z_s}\gsim350\GeV$, the branching to $f\bar{f}$ drops below 10\%. However, it is the dominant annihilation channel for $m_{Z_s}\lsim240\GeV$.

In order to get the observed relic abundance the total annihilation cross section must be $\mathcal{O}(1)$ pb i.e. $\langle\sigma v\rangle = \langle\sigma v\rangle_{Z^0 h^0} + \langle\sigma v\rangle_{f\bar{f}}\approx 1$ pb.  The right-hand plot of Fig.\ref{dmplot} displays the values of $g_s$ vs. $m_{\tilde{W}_s}$ for which the correct relic abundance is obtained, fixing $m_{Z_s}=600~\GeV$ and $\tan\beta=4$. 

Alternatively, if we insist on naturalness and still wish to preserve the global $U(1)_s$ symmetry, we can give up on having the LSiP as dark matter and instead interpret the direct detection bounds as a constraint on the LSiP relic abundance. In that case the bound can be re-cast as:
\begin{equation}
\kappa^2_{\text{LSiP}-Z}\frac{\rho}{\rho_0}\lsim(3.7\times10^{-4})^2,
\end{equation}
where $\rho$ is the local dark matter density and $\rho_0\sim0.3~\GeV/\text{cm}^3$. For an LSiP with electroweak charge such as $\tilde{\Sigma}_s$, that implies:
\begin{equation}
\rho\lsim 10^{-6}~\rho_0.
\end{equation}
Achieving such an under-abundant relic density is challenging. As mentioned before, the annihilation cross section can be significantly increased by allowing the LSiP to go to on-shell states that are charged under $SU(2)_s$, such as $\phi$'s or G-quarks. Even so, the increase is cross section relative to the $Z^0h^0$ channel scales as $\cos^{-4}\beta$, a gain of $\approx3\times10^2$ over what is shown in Fig.\ref{dmplot}.  Although an additional boost may be achieved with large values of $g_s$, naively resonant annihilation through $Z_s$ would still be necessary in order to deplete the LSiP by six orders of magnitude relative to $\Omega_\text{DM}$. A tension then appears since the $Z_s$ width is also increased to $\Gamma_{Z_s}\sim\OO(1-10)~\GeV$, intrinsically limiting the efficiency of resonant annihilation. Other channels such as t-channel annihilation into $Z_s Z_s$ cannot easily achieve the required under-adundance either, even for $\alpha_s \approx 1$. Note that need for $\Omega_{\text{LSiP}}\lsim 10^{-6}~\Omega_\text{DM}$ only holds if the LSiP is dominantly the $SU(2)_\text{EW}$ doublet $\tilde{\Sigma}_s$. Hence, stable Dirac LSiPs that are dominantly $\tilde\Phi_s$ or $\tilde{W}_s$ may still be viable.  

In summary, there seems to be a basic tension to achieving a combination of a) having an intact $U(1)_s$, b) having a stable Dirac particle charged under $U(1)_s$ and $SU(2)_\text{EW}$, and c) naturalness. The easiest ways out are to break the $U(1)_s$ completely or to a $Z_2$ (in both cases the LSP will be Majorana) and making the spin-independent cross section inelastic, or to preserve the $U(1)_s$ but identify it with an existing SM charge such as $U(1)_B$ or $U(1)_L$, in which case any heavy particles could decay to SM quarks or leptons, so that direct detection constraints will not apply.

\subsection{Bosonic LSiP Dark Matter}
So far we have focused on fermionic LSiP dark matter, but it is worth commenting on bosonic LSiP dark matter as well. Many of the same issues will arise here, again due to $Z-Z'$ mixing.

The candidates for a bosonic LSiP WIMP are the $^{(s)}$ charged fields, i.e., $\phi^{(s)}, \bar{\phi}^{(s)}, \sigma_d^{(s)}$ and $\sigma_u^{(s)}$ among the scalars and $W_\mu^{(s)}$ among the vectors. Examining the spectrum of Figure \ref{fig:Sspectrum} we see an immediate tension in our simplified limit. Although the usual $SU(2)$ D-term pushes $\sigma_0^{(s)}$ down in mass, the NMSSM-like term raises the electrically neutral component in general above the electrically charged component. Thus, even to find a reasonable DM candidate we must explore a broader range of parameter space.

Even so, for a suitable candidate, because it is charged under $SU(2)_s$, the $Z-Z'$ mixing will induce the same large $Z$-exchange direct detection process, meaning that only a very light WIMP would be allowed, or under a situation where $U(1)_s$ is broken down to $Z_2^s$. 

An interesting possibility in the broader space would be if the $W^{(s)}$ was the LSiP, yielding an interesting possibility of vector DM. It, too, would interact through the $Z$, and so similar model building constraints apply.

Clearly the model space is broad, and is beyond the scope of what we can accomplish here. We have only begun the discussion of DM in Sister Higgs models.

\section{Collider Implications of an LS$\textsc{i}$P LSP}

\begin{figure}[b] 
\includegraphics[width=3in]{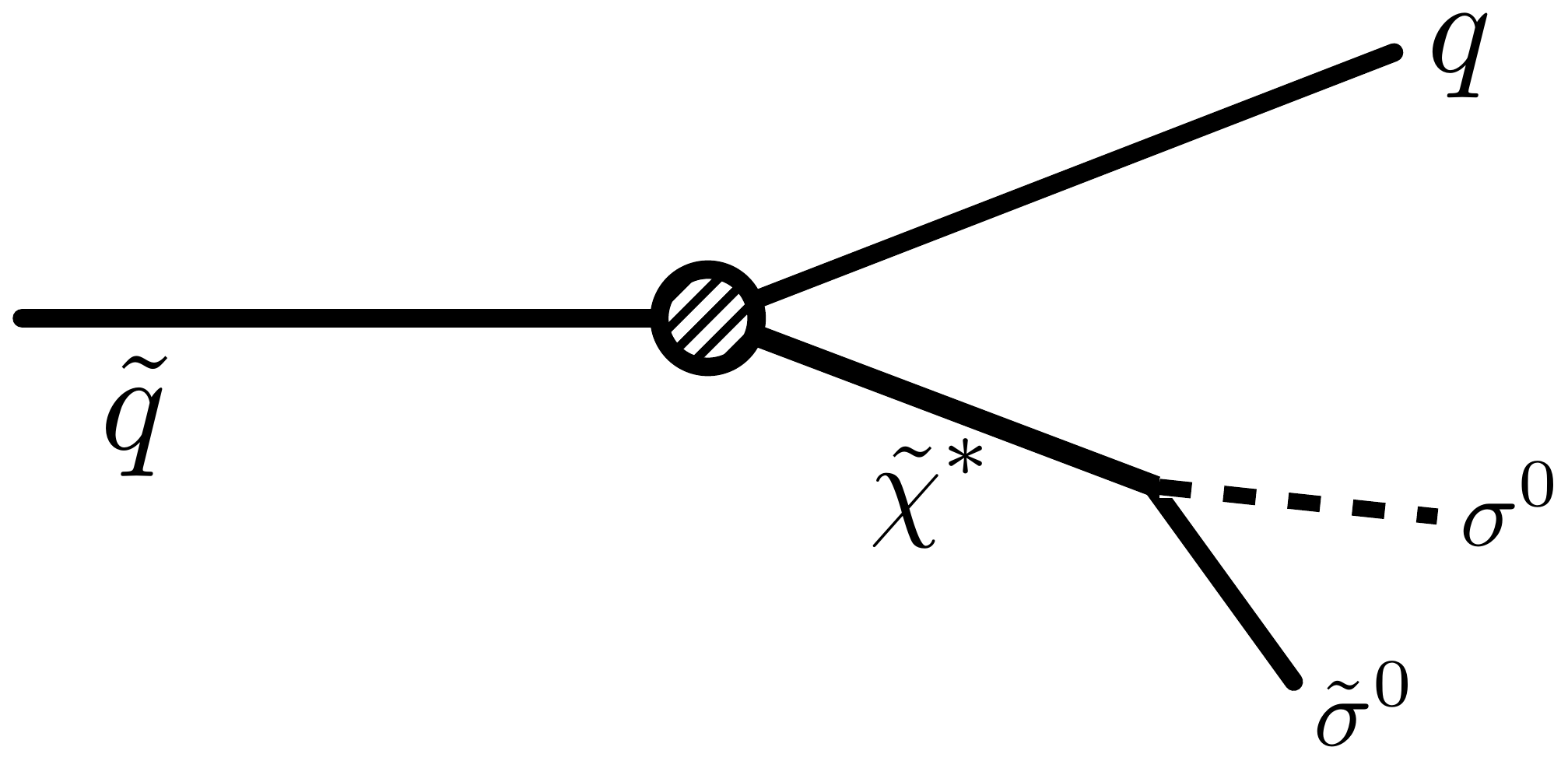}
\caption{Topology of a 3-body squark decay to two invisible sister particles.}\label{squark3body}
\end{figure}

If the LSP carries sister charge, the collider phenomenology of pair produced superpartners is modified in an interesting way. Fig.\ref{squark3body} illustrates that for a squark decay. Since the squark does not carry sister-charge, it can only decay to final states with pairs of sister-charged particles. Spin-statistics requires that the final state contains the lightest sister fermion and the lightest sister boson (note that both are stable). If such decay is mediated by an off-shell neutralino, its kinematics will be that of a 3-body decay, where two of the three final states are ``invisible".

An analogous situation occurs in the MSSM with an off-shell Bino mediating a 3-body decay of the squark to a quark, a neutrino and a sneutrino. Generically a sneutrino LSP would be severely ruled out by direct detection; however those bounds could be evaded in scenarios with mixed-sneutrino LSP \cite{ArkaniHamed:2000bq}, or a sneutrino NLSP which subsequently decays to a neutrino and a gravitino.

\begin{figure}[t] 
\includegraphics[width=6.5in]{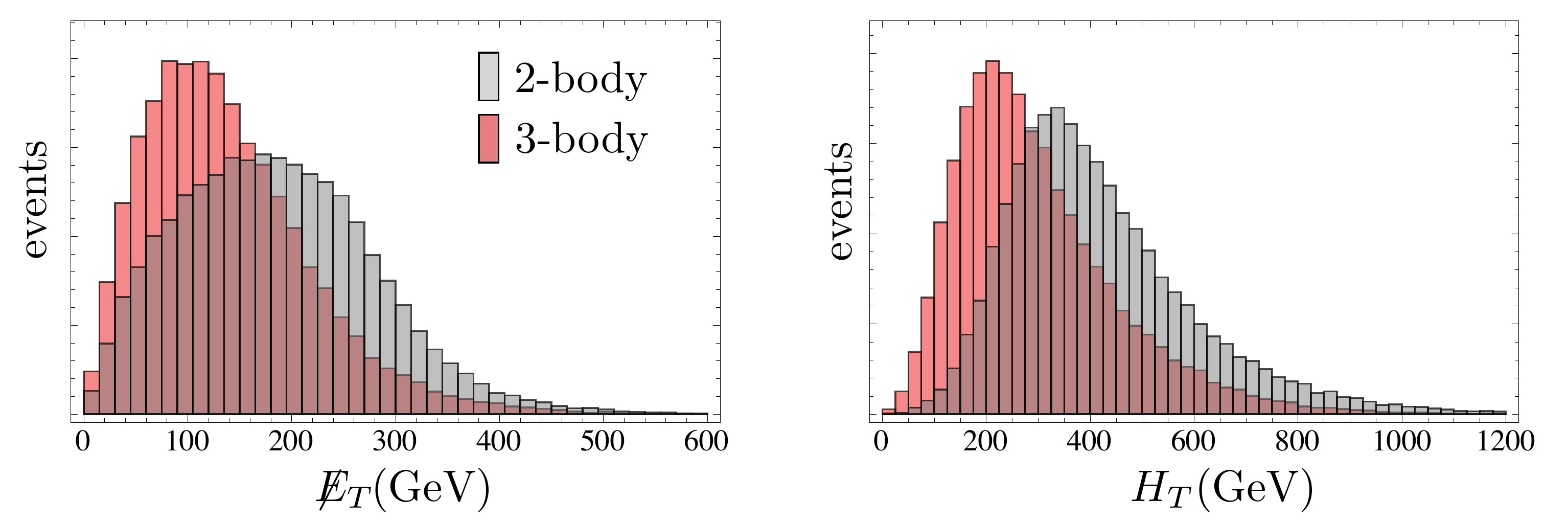}
\caption{Contrast between the $\met$ and $H_T$ distributions of 2- and 3-body squark decays. We assume $m_{\tilde{q}}=400~\GeV$; for the 2-body decay, the mass of the invisible particle is 200 GeV, and for the 3-body decay the two invisible particles have mass of 100 GeV.}\label{METHT}
\end{figure}

An interesting implication of the 3-body kinematics for collider signatures of squarks is a reduction in the visible and missing energy of the events, decreasing the efficiency of these types of signals to pass standard cuts and making its discovery more challenging. Fig.\ref{METHT} contrasts the distributions of missing transverse energy and $H_T$\footnote{$H_T$ is defined as the scalar sum of the $p_T$ of all jets in the event.} for the two types of decay, assuming a squark mass $m_{\tilde{q}}=400~\GeV$. For the 2-body topology, the neutralino mass is chosen to be $m_{\tilde{\chi}^0}=200~\GeV$, and for the 3-body topology the two invisible final state particles have masses $m_{\sigma^0}=m_{\tilde{\sigma}^0}=100~\GeV$.

As a brief illustration of the effects of this different kinematics, we perform a Monte Carlo simulation\footnote{We use $\mathtt{MadGraph5}$ for event simulation, $\mathtt{Pythia6}$ for decays, showering and hadronization and $\mathtt{PGS4}$ for detector response.} of the two benchmark scenarios introduced in Fig.\ref{METHT}, and compare their relative efficiency to pass a standard jets plus $\met$ search. Specifically, we consider the baseline selection in the multijets plus missing transverse energy search performed by CMS \cite{CMSSUS12011} with 4.98 $\text{fb}^{-1}$. Such selection required events with $H_T>500~\GeV$, $\mbox{$H_T\hspace{-0.27in}\not\hspace{0.18in}$}>200~\GeV$, and at least 3 central jets with $p_T>50~\GeV$, among other requirements.
Assuming first and second generation degenerate squarks, this CMS search constrained the production cross section of the 2-body benchmark to be $\sigma_{\tilde{q}\tilde{q}^*}\lsim1.2$~pb.
This cross section is roughly half of the reference QCD cross section at NLO, and therefore the 2-body benchmark is ruled out. The efficiency for the 3-body topology
is roughly 30\% of its 2-body counterpart. Naively applying this lowered efficiency to the cross section limit, we get a bound $\sigma_{\tilde{q}\tilde{q}^*}\lsim4$~pb, suggesting that the 3-body topology is slightly beyond this search's reach.

Dedicated searches for more challenging SUSY signatures, such as stop and sbottom pair production are also impacted by the 3-body decay kinematics. We briefly look into one more case, namely, sbottom searches targeting the 2-body topology $\tilde{b}_1\rightarrow b\tilde{\chi}^0$. ATLAS performed a sbottom dedicated search \cite{Aad:2011cw} with 2~$\text{fb}^{-1}$ of integrated luminosity. The baseline event selection required $\met>130~\GeV$ and two b-tagged jets with $p_T^{j_1}>130~\GeV$ and $p_T^{j_2}>50~\GeV$, among other requirements. Assuming a sbottom mass $m_{\tilde{b}_1}=300~\GeV$, we compared the selection efficiencies for the 2- and 3-body topologies, assuming $50~\GeV$ for the mass of the invisible particle in the 2-body decay, and $25~\GeV$ for both invisible particles in the 3-body decay. The efficiency of the 3-body topology relative to the 2-body topology is $\sim33\%$. Additional signal regions were defined by placing cuts on the boost-corrected contransverse mass \cite{Polesello:2009rn}, $m_{CT}$, which have even lower relative efficiencies. The ATLAS search excluded this 2-body decay mass point, bounding the sbottom production cross section to be $\sigma_{\tilde{b}_1\tilde{b}_1^*}\lsim0.44$~pb. Conservatively using the relative efficiency of the baseline selection to infer the bound on the 3-body topology, we obtain $\sigma_{\tilde{b}_1\tilde{b}_1^*}\lsim1.3$~pb, which is at the borderline of ATLAS' exclusion reach.

While we have only touched on the phenomenology of such decays, we can see that there are clearly interesting results: squarks of 400 GeV and sbottoms of 300 GeV may plausibly be allowed under current LHC limits. The impact of 3-body topologies in LHC searches is a rich subject that deserves more careful investigation. We defer this to future work.

\section{The Collider Physics of G-Quarks}
The phenomenology of the G-quarks is intimately tied into a number of questions: first, are we considering a scenario in which there is a preserved sister charge or not? If there is a sister charge, is it identified with any of the existing global standard model quantum numbers (i.e., baryon or lepton number)? Moreover, if there is a sister charge, and R-parity is conserved, is the lightest sister-charged particle (LSiP) also the lightest supersymmetric particle (LSP)?

Let us begin with the simplest case - there is a single vectorlike pair of G-quarks $D_g$ and $D_g^c$. These fields can interact with the standard model via $\bar \Phi D_g D^c$ which preserves a $U(1)_s$. We could have instead included $ \Phi D_g D^c+\Sigma_d D_g^c Q$, which gives opposite $U(1)_s$ assignments, but since $D_g$ falls into a $5$ with $\Sigma_d$, it is natural to give them the same charge assignments. 

If this sister charge is broken, it can be broken spontaneously or explicitly. If broken spontaneously by $\phi$ and $\sigma$ vevs, it should be gauged to avoid the presence of a dangerous goldstone boson. In this case the production of G-quarks will naturally yield $D_g^{(c)}\rightarrow D^c \phi (\sigma)$. With $\phi (\sigma) \rightarrow \bar b b$ we can naturally expect six-jet events with three-jet resonances. CMS has performed a search for three-jet resonances in the context of hadronic RPV-gluino decays \cite{CMS-EXO-11-060}, excluding gluino masses $m_{\tilde{g}}\lsim460~\GeV$. The limits from this search on G-quarks are much weaker, since the $D_g\bar{D}_g$ production cross section is roughly an order of magnitude smaller than gluino pair-production.  A more quantitative study of the limits on such objects can be found in \cite{crpv}.

The G-squarks will mix with SM squarks, and so if they are lighter than the G-quarks, they will produce conventional jet+$\met$ signals when pair-produced. However, if heavier than the G-quarks, and assuming a conventional LSP such as a Bino, they will decay $\tilde d_g \rightarrow D_g + \chi^0 \rightarrow j \bar b b\tilde \chi^0$. Thus, they will produce six jet + $\met$ events, with large numbers of b-tags. For large G-quark -G-squark splittings, standard jets+$\met$ searches might be sensitive to these signatures, although with smaller reach relative to MSSM squarks due to the reduced G-squark cross section. (The limits on MSSM squarks usually quoted by LHC searches implicitly assume degenerate 1st and 2nd generation squarks, with an effective cross section eight times larger than the cross section of a single squark species). On the other hand, for more compressed spectra the missing energy in the events may be significantly reduced, rendering high-$\met$ searches insensitive to such models. However, searches with high-jet multiplicity or/and high b-jet multiplicity may be effective for these topologies. Still, the cross section may be the limiting factor in sensitivity to G-quarks.

\begin{figure}[t] 
\includegraphics[width=5.7in]{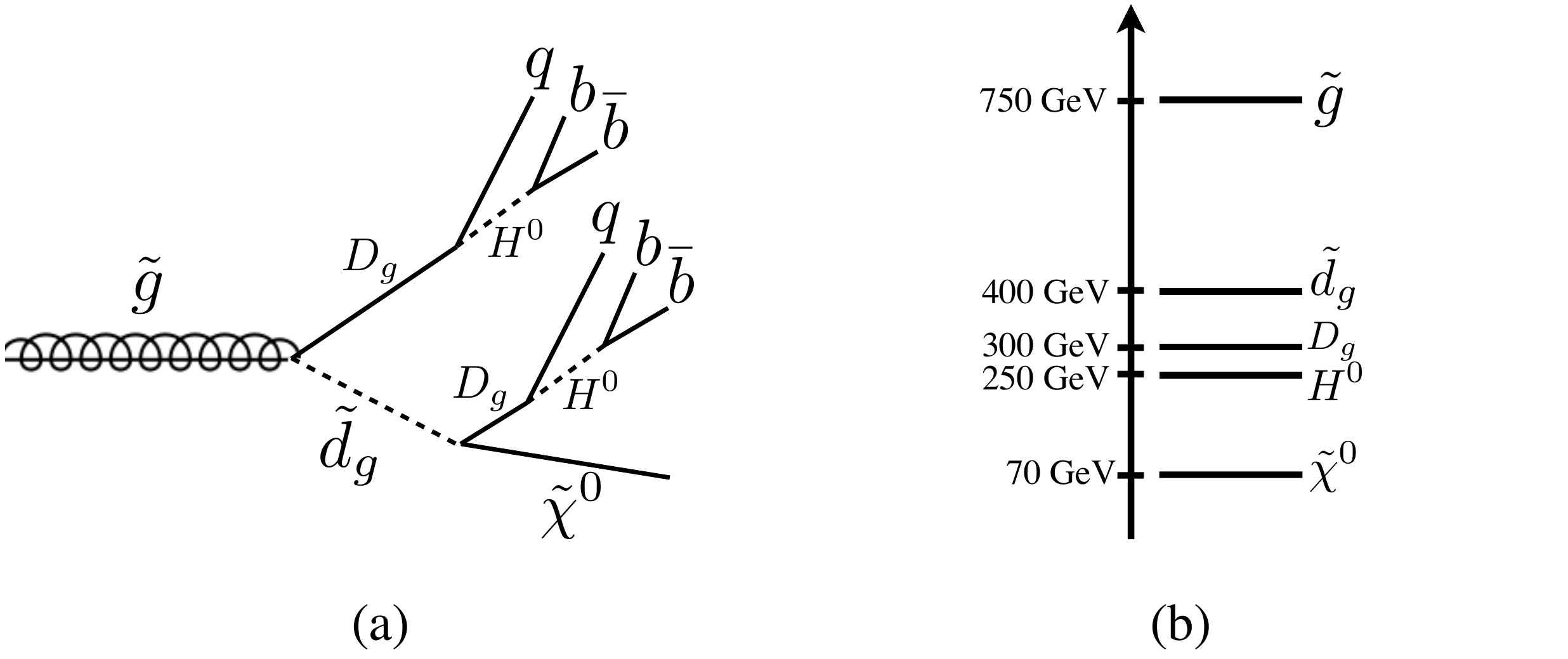}
\caption{(a) Gluino decay topology to $\tilde{d}_g D_g$, resulting in final states with $2j + 4b + \tilde\chi^0$. (b) Benchmark spectrum with small splitting between $m_{\tilde{d}_g}$ and $m_{D_g}+m_{\tilde\chi^0}$.}\label{goGG}
\end{figure}

An interesting scenario to consider is the case of gluino-boosted G-(s)quark production. If the G-(s)quarks are lighter than the gluinos with all other squarks heavy, the on-shell decay $\tilde g \rightarrow D_g \tilde d_g$ might dominate. This decay topology is illustrated in Fig.\ref{goGG}(a) - each gluino decays to $2j + 4b + \tilde\chi^0$ final states, for a total of 12 jet + $\met$ events (with large numbers of b-tags). Dedicated searches with high b-jet multiplicity could be sensitive to such final states, such as the search performed by ATLAS in missing energy and at least 3 b-jets \cite{ATLAS-CONF-2012-058}. This search, however, places a strong requirement on missing transverse energy ($\met>160~\GeV$), which narrows its sensitivity to spectra with large mass splittings between G-quarks and G-squarks. In Fig.\ref{goGG}(b) we illustrate a typical spectrum which the ATLAS 3-bjets+$\met$ search is not sensitive to. In the rest frame of ${\tilde{d}_g}$, the 3-momentum of its neutralino daughter $\tilde{\chi}^0$ is $|\vec{p}_{\tilde\chi^0}|\sim62~\GeV$, and hence this signal has a very low efficiency to pass the $\met>160~\GeV$ requirement of this search. Despite its low missing energy, this particular spectrum is distinctive enough to be caught by high jet-multiplicity searches, and indeed, through our MC simulations we estimate that this mass spectrum is excluded by the ATLAS search in this channel\cite{ATLAS-CONF-2012-037}.  This search looks for events with up to 9 hard jets, and bounds the gluino cross section in the benchmark of Fig.\ref{goGG} to be $\sigma_{\tilde{g}\tilde{g}}\lsim0.1~\text{pb}$, which is roughly the value of the reference QCD cross section for a 750 GeV gluino. Hence, this point is marginally excluded.

If this sister charge is not broken by $\phi$ and $\Sigma$ vevs, there must be a larger gauged sister group, i.e., $SU(2)_s^{\rm gauge} \times U(1)_\Sigma^{\rm global}$, which leaves some residual $U(1)_{s}$. In this case, we must determine the identity of the LSiP. Assuming it is stable, it is most easily neutral, such as the $\phi$, in which case the decay $D_g \rightarrow D + {\rm LSiP}$ would resemble a standard SUSY squark search. Likewise $\tilde d_g \rightarrow D_g \tilde \chi^0 \rightarrow D + \tilde\chi^0 + {\rm LSiP}$ could be discoverable as well with enough luminosity.

The sister charge can be broken explicitly if the sister group is $SU(2)_s^{\rm gauge} \times U(1)_\Sigma^{\rm global}$. I.e., we can write $U(1)_\Sigma$ breaking terms $\bar \phi D_g D^c +\phi D_g D^c+\Sigma_d D_g^c Q$. In this case, both G-quarks decay is as in the broken $U(1)$ case above, with multijets and many b-tags.

Another interesting case is when the $U(1)_s$ is identified with some SM global charge. As we discuss in the appendix, the operator $U D_s^c D_s^c$ naturally identifies $U(1)_s$ with baryon number, but does not break R-parity. In this case, while the G-quarks without sister charge would decay into three jets, the sister-charged G-quark would decay $\tilde  D_s^c \rightarrow \tilde u D_s\rightarrow U + D +b \bar b + \chi_0$, yielding eight-jet plus $\met$ signals, and sister charged G-squarks would decay $\tilde d_g^{(s)} \rightarrow u D_g^c \rightarrow u d \bar b b$, i.e., we would have eight jet events, with four jet subresonances. (Here, the G-squarks of different sister charge have different R-parity under the intact R-parity after $\phi, \Sigma$ acquire vevs.)

In summary, the presence of G-quarks can radically alter our expectations for phenomenology. In particular, if the G-quarks are lighter than the gluinos, it can dramatically enhance the number of jets (in particular b-jets) in the event, and suppress the $\met$ signal.

\section{Conclusions}\label{sec:conclusions}
If there is new physics at the weak scale, the Higgs boson and its properties are the first indication of what that may be. The value of 125 GeV constrains a wide range of supersymmetric models, and may be pointing us to a new contribution to the Higgs quartic. Models such as the NMSSM can raise it, but at the cost of including pure singlets, and only with very large couplings.

In the context of the NMSSM, however, it is possible to replace $H_d$ with a new field $\Sigma_d$, a ``Sister Higgs,'' and still yield the usual tree-level corrections to the Higgs mass. Such a field need not couple to ordinary matter, and, indeed, would not if it carries a charge under some new sister group $G_s$. If that group is gauged, new contributions to the NMSSM-like superpotential term $\Phi H_u \Sigma_d$ can drive it large, easily boosting the Higgs mass up to 125 GeV, or beyond.

The presence of this new Higgs has a number of consequences: as with a Type I 2HDM, a light charged Higgs is now possible. The overall consequences allow a modification to the Higgs properties, such as a boosted VBF $\gamma\gamma$ signal, as well as a boosted inclusive $\gamma\gamma$ signal, with suppressed decays to fermions \cite{SHsignals}.

Once we have expanded the theory to include the sister Higgs, we could also expect yet more Higgses when $G_s$ is non-Abelian. With these fields we can write down the R-parity violating operator $H_d \Sigma_d E^c$ once $G_s$ is broken. Because this operator couples only to right handed fermions, it is far less constrained that traditional lepton-number violating RPV operators. While it changes the phenomenology of SUSY decays, if the finals state contains $\tau$'s, the current search strategies should find it. At the same time, it motivates searches for leptoquarks, with squarks decaying to $j \ell^\pm$, but importantly without an associated missing energy signal (as there is no accompanying channel with a neutrino).

The $Z'$ in these models is natural, and can be quite light without significant constraints from the $\rho$ parameter or searches for dilepton resonances. Corrections to $Zh$ production can be sizeable at the Tevatron but are not expected to be large at the LHC without additional couplings to SM fermions (although such corrections could come from mixing with the G-quarks at the cost of flavor constraints). 

Appeals to grand-unification motivate us to consider the presence of colored fields which are charged under $G_s$ as well. Unlike conventional fourth-generations, these vectorlike quarks decay dominantly through $\phi$ emission (or $\phi-h$ mixing induced $h$ emission). These G-quarks can have important effects on phenomenology. In particular, if they are lighter than the gluino, they can open up new decay channels with lower missing energy and more jets than conventional gluino decays. The phenomenology of G-quarks (and G-squarks) is very rich and we have only begun the discussion in this work.

In the presence of a sister charge, the lightest sister-charged particle (LSiP) would be stable. If it is a Dirac fermion, constraints from direct detection searches would have excluded it in all but the narrowest corners of parameter space where the WIMP is very under abundant or is very light. It may thus be most natural that symmetry should be broken, at least to a $Z_2$, or become identified with $B$ or $L$ so that the LSiP can decay into it.

Regardless, if the LSP also carries sister charge, squarks may only decay via three-body decays. Such decays have reduced missing energy and $H_T$, and are consequently less constrained by existing jets+$\met$ searches.

In summary - very complex phenomenology can arise by the simple extension of the MSSM by the presence of $G_s$-charged sister-Higgs fields and their related G-quarks. A low energy theory that is not the MSSM, nor even the NMSSM, but one expanded with additional Higgs fields easily provides for a Higgs mass as large as 125 GeV and a profoundly changed LHC phenomenology. As more data accumulate, we shall see soon if such a rich Higgs sector is realized in nature.

\section*{Acknowledgements}
We thank N.Arkani-Hamed and I. Yavin for helpful conversations.  We thank Natalia Toro for showing us it takes GUTs to think about conformality. NW thanks G.~Vukmirovic for her support.  NW is supported by NSF grant \#0947827.  Fermilab is operated by Fermi Research Alliance, LLC, under Contract DE-AC02-07CH11359 with the United States Department of Energy.

\bibliography{sisterhiggs}

\begin{thebibliography}{53}
\expandafter\ifx\csname natexlab\endcsname\relax\def\natexlab#1{#1}\fi
\expandafter\ifx\csname bibnamefont\endcsname\relax
  \def\bibnamefont#1{#1}\fi
\expandafter\ifx\csname bibfnamefont\endcsname\relax
  \def\bibfnamefont#1{#1}\fi
\expandafter\ifx\csname citenamefont\endcsname\relax
  \def\citenamefont#1{#1}\fi
\expandafter\ifx\csname url\endcsname\relax
  \def\url#1{\texttt{#1}}\fi
\expandafter\ifx\csname urlprefix\endcsname\relax\def\urlprefix{URL }\fi
\providecommand{\bibinfo}[2]{#2}
\providecommand{\eprint}[2][]{\url{#2}}

\bibitem[{\citenamefont{Incandela}(2012)}]{cmshiggstalk}
\bibinfo{author}{\bibfnamefont{J.}~\bibnamefont{Incandela}},
  \emph{\bibinfo{title}{{Update on the Standard Model Higgs searches in CMS}}}
  (\bibinfo{year}{2012}).

\bibitem[{\citenamefont{Gianotti}(2012)}]{atlashiggstalk}
\bibinfo{author}{\bibfnamefont{F.}~\bibnamefont{Gianotti}},
  \emph{\bibinfo{title}{{Update on the Standard Model Higgs searches in
  ATLAS}}} (\bibinfo{year}{2012}).

\bibitem[{\citenamefont{Hall et~al.}(2011)\citenamefont{Hall, Pinner, and
  Ruderman}}]{Hall:2011aa}
\bibinfo{author}{\bibfnamefont{L.~J.} \bibnamefont{Hall}},
  \bibinfo{author}{\bibfnamefont{D.}~\bibnamefont{Pinner}}, \bibnamefont{and}
  \bibinfo{author}{\bibfnamefont{J.~T.} \bibnamefont{Ruderman}}
  (\bibinfo{year}{2011}), \eprint{1112.2703}.

\bibitem[{\citenamefont{Harnik et~al.}(2004)\citenamefont{Harnik, Kribs,
  Larson, and Murayama}}]{Harnik:2003rs}
\bibinfo{author}{\bibfnamefont{R.}~\bibnamefont{Harnik}},
  \bibinfo{author}{\bibfnamefont{G.~D.} \bibnamefont{Kribs}},
  \bibinfo{author}{\bibfnamefont{D.~T.} \bibnamefont{Larson}},
  \bibnamefont{and} \bibinfo{author}{\bibfnamefont{H.}~\bibnamefont{Murayama}},
  \bibinfo{journal}{Phys.Rev.} \textbf{\bibinfo{volume}{D70}},
  \bibinfo{pages}{015002} (\bibinfo{year}{2004}), \eprint{hep-ph/0311349}.

\bibitem[{\citenamefont{Chang et~al.}(2005)\citenamefont{Chang, Kilic, and
  Mahbubani}}]{Chang:2004db}
\bibinfo{author}{\bibfnamefont{S.}~\bibnamefont{Chang}},
  \bibinfo{author}{\bibfnamefont{C.}~\bibnamefont{Kilic}}, \bibnamefont{and}
  \bibinfo{author}{\bibfnamefont{R.}~\bibnamefont{Mahbubani}},
  \bibinfo{journal}{Phys.Rev.} \textbf{\bibinfo{volume}{D71}},
  \bibinfo{pages}{015003} (\bibinfo{year}{2005}), \eprint{hep-ph/0405267}.

\bibitem[{\citenamefont{Evans et~al.}(2012)\citenamefont{Evans, Ibe, and
  Yanagida}}]{Evans:2012uf}
\bibinfo{author}{\bibfnamefont{J.~L.} \bibnamefont{Evans}},
  \bibinfo{author}{\bibfnamefont{M.}~\bibnamefont{Ibe}}, \bibnamefont{and}
  \bibinfo{author}{\bibfnamefont{T.~T.} \bibnamefont{Yanagida}}
  (\bibinfo{year}{2012}), \eprint{1204.6085}.

\bibitem[{\citenamefont{Espinosa and Quiros}(1993)}]{Espinosa:1992hp}
\bibinfo{author}{\bibfnamefont{J.}~\bibnamefont{Espinosa}} \bibnamefont{and}
  \bibinfo{author}{\bibfnamefont{M.}~\bibnamefont{Quiros}},
  \bibinfo{journal}{Phys.Lett.} \textbf{\bibinfo{volume}{B302}},
  \bibinfo{pages}{51} (\bibinfo{year}{1993}), \eprint{hep-ph/9212305}.

\bibitem[{\citenamefont{Masip et~al.}(1998)\citenamefont{Masip, Munoz-Tapia,
  and Pomarol}}]{Masip:1998jc}
\bibinfo{author}{\bibfnamefont{M.}~\bibnamefont{Masip}},
  \bibinfo{author}{\bibfnamefont{R.}~\bibnamefont{Munoz-Tapia}},
  \bibnamefont{and} \bibinfo{author}{\bibfnamefont{A.}~\bibnamefont{Pomarol}},
  \bibinfo{journal}{Phys.Rev.} \textbf{\bibinfo{volume}{D57}},
  \bibinfo{pages}{R5340} (\bibinfo{year}{1998}), \eprint{hep-ph/9801437}.

\bibitem[{\citenamefont{Batra et~al.}(2004{\natexlab{a}})\citenamefont{Batra,
  Delgado, Kaplan, and Tait}}]{Batra:2003nj}
\bibinfo{author}{\bibfnamefont{P.}~\bibnamefont{Batra}},
  \bibinfo{author}{\bibfnamefont{A.}~\bibnamefont{Delgado}},
  \bibinfo{author}{\bibfnamefont{D.~E.} \bibnamefont{Kaplan}},
  \bibnamefont{and} \bibinfo{author}{\bibfnamefont{T.~M.~P.}
  \bibnamefont{Tait}}, \bibinfo{journal}{JHEP} \textbf{\bibinfo{volume}{02}},
  \bibinfo{pages}{043} (\bibinfo{year}{2004}{\natexlab{a}}),
  \eprint{hep-ph/0309149}.

\bibitem[{\citenamefont{Batra et~al.}(2004{\natexlab{b}})\citenamefont{Batra,
  Delgado, Kaplan, and Tait}}]{Batra:2004vc}
\bibinfo{author}{\bibfnamefont{P.}~\bibnamefont{Batra}},
  \bibinfo{author}{\bibfnamefont{A.}~\bibnamefont{Delgado}},
  \bibinfo{author}{\bibfnamefont{D.~E.} \bibnamefont{Kaplan}},
  \bibnamefont{and} \bibinfo{author}{\bibfnamefont{T.~M.~P.}
  \bibnamefont{Tait}}, \bibinfo{journal}{JHEP} \textbf{\bibinfo{volume}{06}},
  \bibinfo{pages}{032} (\bibinfo{year}{2004}{\natexlab{b}}),
  \eprint{hep-ph/0404251}.

\bibitem[{\citenamefont{Kribs et~al.}(2008)\citenamefont{Kribs, Poppitz, and
  Weiner}}]{Kribs:2007ac}
\bibinfo{author}{\bibfnamefont{G.~D.} \bibnamefont{Kribs}},
  \bibinfo{author}{\bibfnamefont{E.}~\bibnamefont{Poppitz}}, \bibnamefont{and}
  \bibinfo{author}{\bibfnamefont{N.}~\bibnamefont{Weiner}},
  \bibinfo{journal}{Phys.Rev.} \textbf{\bibinfo{volume}{D78}},
  \bibinfo{pages}{055010} (\bibinfo{year}{2008}), \eprint{0712.2039}.

\bibitem[{\citenamefont{Giudice and Strumia}(2012)}]{Giudice:2011cg}
\bibinfo{author}{\bibfnamefont{G.~F.} \bibnamefont{Giudice}} \bibnamefont{and}
  \bibinfo{author}{\bibfnamefont{A.}~\bibnamefont{Strumia}},
  \bibinfo{journal}{Nucl. Phys.} \textbf{\bibinfo{volume}{B858}},
  \bibinfo{pages}{63} (\bibinfo{year}{2012}), \eprint{1108.6077}.

\bibitem[{\citenamefont{Espinosa and Quiros}(1998)}]{Espinosa:1998re}
\bibinfo{author}{\bibfnamefont{J.~R.} \bibnamefont{Espinosa}} \bibnamefont{and}
  \bibinfo{author}{\bibfnamefont{M.}~\bibnamefont{Quiros}},
  \bibinfo{journal}{Phys. Rev. Lett.} \textbf{\bibinfo{volume}{81}},
  \bibinfo{pages}{516} (\bibinfo{year}{1998}), \eprint{hep-ph/9804235}.

\bibitem[{\citenamefont{Barger et~al.}(1990)\citenamefont{Barger, Hewett, and
  Phillips}}]{Barger:1989fj}
\bibinfo{author}{\bibfnamefont{V.~D.} \bibnamefont{Barger}},
  \bibinfo{author}{\bibfnamefont{J.}~\bibnamefont{Hewett}}, \bibnamefont{and}
  \bibinfo{author}{\bibfnamefont{R.}~\bibnamefont{Phillips}},
  \bibinfo{journal}{Phys.Rev.} \textbf{\bibinfo{volume}{D41}},
  \bibinfo{pages}{3421} (\bibinfo{year}{1990}).

\bibitem[{\citenamefont{Staub}(2008)}]{Staub:2008uz}
\bibinfo{author}{\bibfnamefont{F.}~\bibnamefont{Staub}} (\bibinfo{year}{2008}),
  \eprint{0806.0538}.

\bibitem[{\citenamefont{Staub}(2011)}]{Staub:2010jh}
\bibinfo{author}{\bibfnamefont{F.}~\bibnamefont{Staub}},
  \bibinfo{journal}{Comput. Phys. Commun.} \textbf{\bibinfo{volume}{182}},
  \bibinfo{pages}{808} (\bibinfo{year}{2011}), \eprint{1002.0840}.

\bibitem[{\citenamefont{Xing et~al.}(2008)\citenamefont{Xing, Zhang, and
  Zhou}}]{Xing:2007fb}
\bibinfo{author}{\bibfnamefont{Z.-z.} \bibnamefont{Xing}},
  \bibinfo{author}{\bibfnamefont{H.}~\bibnamefont{Zhang}}, \bibnamefont{and}
  \bibinfo{author}{\bibfnamefont{S.}~\bibnamefont{Zhou}},
  \bibinfo{journal}{Phys. Rev.} \textbf{\bibinfo{volume}{D77}},
  \bibinfo{pages}{113016} (\bibinfo{year}{2008}), \eprint{0712.1419}.

\bibitem[{\citenamefont{Mahmoudi and Stal}(2010)}]{Mahmoudi:2009zx}
\bibinfo{author}{\bibfnamefont{F.}~\bibnamefont{Mahmoudi}} \bibnamefont{and}
  \bibinfo{author}{\bibfnamefont{O.}~\bibnamefont{Stal}},
  \bibinfo{journal}{Phys.Rev.} \textbf{\bibinfo{volume}{D81}},
  \bibinfo{pages}{035016} (\bibinfo{year}{2010}), \eprint{0907.1791}.

\bibitem[{\citenamefont{{ATLAS
  Collaboration}}(2012{\natexlab{a}})}]{ATLAS-CONF-2012-007}
\bibinfo{author}{\bibnamefont{{ATLAS Collaboration}}}, \bibinfo{type}{Tech.
  Rep.} \bibinfo{number}{ATLAS-CONF-2012-007}, \bibinfo{institution}{CERN},
  \bibinfo{address}{Geneva} (\bibinfo{year}{2012}{\natexlab{a}}).

\bibitem[{\citenamefont{Fan et~al.}(2011)\citenamefont{Fan, Krohn, Langacker,
  and Yavin}}]{Fan:2011vw}
\bibinfo{author}{\bibfnamefont{J.}~\bibnamefont{Fan}},
  \bibinfo{author}{\bibfnamefont{D.}~\bibnamefont{Krohn}},
  \bibinfo{author}{\bibfnamefont{P.}~\bibnamefont{Langacker}},
  \bibnamefont{and} \bibinfo{author}{\bibfnamefont{I.}~\bibnamefont{Yavin}},
  \bibinfo{journal}{Phys.Rev.} \textbf{\bibinfo{volume}{D84}},
  \bibinfo{pages}{105012} (\bibinfo{year}{2011}), \eprint{1106.1682}.

\bibitem[{\citenamefont{Arkani-Hamed et~al.}(2006)\citenamefont{Arkani-Hamed,
  Delgado, and Giudice}}]{ArkaniHamed:2006mb}
\bibinfo{author}{\bibfnamefont{N.}~\bibnamefont{Arkani-Hamed}},
  \bibinfo{author}{\bibfnamefont{A.}~\bibnamefont{Delgado}}, \bibnamefont{and}
  \bibinfo{author}{\bibfnamefont{G.~F.} \bibnamefont{Giudice}},
  \bibinfo{journal}{Nucl. Phys.} \textbf{\bibinfo{volume}{B741}},
  \bibinfo{pages}{108} (\bibinfo{year}{2006}), \eprint{hep-ph/0601041}.

\bibitem[{\citenamefont{Alves et~al.}()\citenamefont{Alves, Fox, , Primulando,
  and Weiner}}]{ustodo}
\bibinfo{author}{\bibfnamefont{D.~S.~M.} \bibnamefont{Alves}},
  \bibinfo{author}{\bibfnamefont{P.~J.} \bibnamefont{Fox}}, ,
  \bibinfo{author}{\bibfnamefont{R.}~\bibnamefont{Primulando}},
  \bibnamefont{and} \bibinfo{author}{\bibfnamefont{N.}~\bibnamefont{Weiner}},
  \bibinfo{note}{{In Preparation}}.

\bibitem[{\citenamefont{Aad et~al.}(2012{\natexlab{a}})}]{Aad:2011ch}
\bibinfo{author}{\bibfnamefont{G.}~\bibnamefont{Aad}} \bibnamefont{et~al.}
  (\bibinfo{collaboration}{ATLAS Collaboration}), \bibinfo{journal}{Phys.Lett.}
  \textbf{\bibinfo{volume}{B709}}, \bibinfo{pages}{158}
  (\bibinfo{year}{2012}{\natexlab{a}}), \eprint{1112.4828}.

\bibitem[{\citenamefont{Aad et~al.}(2012{\natexlab{b}})}]{ATLAS:2012aq}
\bibinfo{author}{\bibfnamefont{G.}~\bibnamefont{Aad}} \bibnamefont{et~al.}
  (\bibinfo{collaboration}{ATLAS Collaboration})
  (\bibinfo{year}{2012}{\natexlab{b}}), \eprint{1203.3172}.

\bibitem[{\citenamefont{{CMS Collaboration}}(2011)}]{CMS-PAS-EXO-11-028}
\bibinfo{author}{\bibnamefont{{CMS Collaboration}}}, \bibinfo{type}{Tech. Rep.}
  \bibinfo{number}{CMS-PAS-EXO-11-028} (\bibinfo{year}{2011}).

\bibitem[{\citenamefont{Aad et~al.}(2012{\natexlab{c}})}]{ATLAS:2012ag}
\bibinfo{author}{\bibfnamefont{G.}~\bibnamefont{Aad}} \bibnamefont{et~al.}
  (\bibinfo{collaboration}{ATLAS Collaboration})
  (\bibinfo{year}{2012}{\natexlab{c}}), \eprint{1203.6580}.

\bibitem[{CMS(2012{\natexlab{a}})}]{CMS-EXO-12-002}
\bibinfo{type}{Tech. Rep.} \bibinfo{number}{EXO-12-002},
  \bibinfo{institution}{CERN}, \bibinfo{address}{Geneva}
  (\bibinfo{year}{2012}{\natexlab{a}}).

\bibitem[{\citenamefont{Aad et~al.}(2012{\natexlab{d}})}]{Aad:2012tj}
\bibinfo{author}{\bibfnamefont{G.}~\bibnamefont{Aad}} \bibnamefont{et~al.}
  (\bibinfo{collaboration}{ATLAS Collaboration})
  (\bibinfo{year}{2012}{\natexlab{d}}), \eprint{1204.2760}.

\bibitem[{\citenamefont{Chatrchyan et~al.}(2012{\natexlab{a}})}]{CMS:2012th}
\bibinfo{author}{\bibfnamefont{S.}~\bibnamefont{Chatrchyan}}
  \bibnamefont{et~al.} (\bibinfo{collaboration}{CMS Collaboration})
  (\bibinfo{year}{2012}{\natexlab{a}}), \eprint{1205.6615}.

\bibitem[{\citenamefont{Chatrchyan
  et~al.}(2012{\natexlab{b}})}]{Chatrchyan:2012sa}
\bibinfo{author}{\bibfnamefont{S.}~\bibnamefont{Chatrchyan}}
  \bibnamefont{et~al.} (\bibinfo{collaboration}{CMS Collaboration})
  (\bibinfo{year}{2012}{\natexlab{b}}), \eprint{1205.3933}.

\bibitem[{\citenamefont{Aad et~al.}(2012{\natexlab{e}})}]{ATLAS:2012ai}
\bibinfo{author}{\bibfnamefont{G.}~\bibnamefont{Aad}} \bibnamefont{et~al.}
  (\bibinfo{collaboration}{ATLAS Collaboration}), \bibinfo{journal}{Phys. Rev.
  Lett.} \textbf{\bibinfo{volume}{108}}, \bibinfo{pages}{241802}
  (\bibinfo{year}{2012}{\natexlab{e}}), \eprint{1203.5763}.

\bibitem[{CMS(2012{\natexlab{b}})}]{CMS-PAS-SUS-12-017}
\bibinfo{type}{Tech. Rep.} \bibinfo{number}{CMS-PAS-SUS-12-017},
  \bibinfo{institution}{CERN}, \bibinfo{address}{Geneva}
  (\bibinfo{year}{2012}{\natexlab{b}}).

\bibitem[{ATL(2012{\natexlab{a}})}]{ATLAS-CONF-2012-069}
\bibinfo{type}{Tech. Rep.} \bibinfo{number}{ATLAS-CONF-2012-069},
  \bibinfo{institution}{CERN}, \bibinfo{address}{Geneva}
  (\bibinfo{year}{2012}{\natexlab{a}}).

\bibitem[{\citenamefont{Chatrchyan
  et~al.}(2012{\natexlab{c}})}]{Chatrchyan:2012ye}
\bibinfo{author}{\bibfnamefont{S.}~\bibnamefont{Chatrchyan}}
  \bibnamefont{et~al.} (\bibinfo{collaboration}{CMS Collaboration})
  (\bibinfo{year}{2012}{\natexlab{c}}), \eprint{1204.5341}.

\bibitem[{ATL(2012{\natexlab{b}})}]{ATLAS-CONF-2012-077}
\bibinfo{type}{Tech. Rep.} \bibinfo{number}{ATLAS-CONF-2012-077},
  \bibinfo{institution}{CERN}, \bibinfo{address}{Geneva}
  (\bibinfo{year}{2012}{\natexlab{b}}).

\bibitem[{ATL(2012{\natexlab{c}})}]{ATLAS-CONF-2012-035}
\bibinfo{type}{Tech. Rep.} \bibinfo{number}{ATLAS-CONF-2012-035},
  \bibinfo{institution}{CERN}, \bibinfo{address}{Geneva}
  (\bibinfo{year}{2012}{\natexlab{c}}).

\bibitem[{ATL(2012{\natexlab{d}})}]{ATLAS-CONF-2012-001}
\bibinfo{type}{Tech. Rep.} \bibinfo{number}{ATLAS-CONF-2012-001},
  \bibinfo{institution}{CERN}, \bibinfo{address}{Geneva}
  (\bibinfo{year}{2012}{\natexlab{d}}).

\bibitem[{ATL(2011{\natexlab{a}})}]{ATLAS-CONF-2011-158}
\bibinfo{type}{Tech. Rep.} \bibinfo{number}{ATLAS-CONF-2011-158},
  \bibinfo{institution}{CERN}, \bibinfo{address}{Geneva}
  (\bibinfo{year}{2011}{\natexlab{a}}).

\bibitem[{ATL(2011{\natexlab{b}})}]{ATLAS-CONF-2011-144}
\bibinfo{type}{Tech. Rep.} \bibinfo{number}{ATLAS-CONF-2011-144},
  \bibinfo{institution}{CERN}, \bibinfo{address}{Geneva}
  (\bibinfo{year}{2011}{\natexlab{b}}).

\bibitem[{\citenamefont{{ATLAS
  Collaboration}}(2012{\natexlab{b}})}]{Collaboration:2012yw}
\bibinfo{author}{\bibnamefont{{ATLAS Collaboration}}}
  (\bibinfo{year}{2012}{\natexlab{b}}), \eprint{1205.0725}.

\bibitem[{\citenamefont{Kribs et~al.}(2009)\citenamefont{Kribs, Martin, and
  Roy}}]{Kribs:2008hq}
\bibinfo{author}{\bibfnamefont{G.~D.} \bibnamefont{Kribs}},
  \bibinfo{author}{\bibfnamefont{A.}~\bibnamefont{Martin}}, \bibnamefont{and}
  \bibinfo{author}{\bibfnamefont{T.~S.} \bibnamefont{Roy}},
  \bibinfo{journal}{JHEP} \textbf{\bibinfo{volume}{01}}, \bibinfo{pages}{023}
  (\bibinfo{year}{2009}), \eprint{0807.4936}.

\bibitem[{ATL(2012{\natexlab{e}})}]{ATLAS-CONF-2012-076}
\bibinfo{type}{Tech. Rep.} \bibinfo{number}{ATLAS-CONF-2012-076},
  \bibinfo{institution}{CERN}, \bibinfo{address}{Geneva}
  (\bibinfo{year}{2012}{\natexlab{e}}).

\bibitem[{\citenamefont{Aprile et~al.}(2011)}]{Aprile:2011hi}
\bibinfo{author}{\bibfnamefont{E.}~\bibnamefont{Aprile}} \bibnamefont{et~al.}
  (\bibinfo{collaboration}{XENON100 Collaboration}),
  \bibinfo{journal}{Phys.Rev.Lett.} \textbf{\bibinfo{volume}{107}},
  \bibinfo{pages}{131302} (\bibinfo{year}{2011}), \eprint{1104.2549}.

\bibitem[{\citenamefont{Tucker-Smith and Weiner}(2001)}]{TuckerSmith:2001hy}
\bibinfo{author}{\bibfnamefont{D.}~\bibnamefont{Tucker-Smith}}
  \bibnamefont{and} \bibinfo{author}{\bibfnamefont{N.}~\bibnamefont{Weiner}},
  \bibinfo{journal}{Phys.Rev.} \textbf{\bibinfo{volume}{D64}},
  \bibinfo{pages}{043502} (\bibinfo{year}{2001}), \eprint{hep-ph/0101138}.

\bibitem[{\citenamefont{Arkani-Hamed et~al.}(2001)\citenamefont{Arkani-Hamed,
  Hall, Murayama, Tucker-Smith, and Weiner}}]{ArkaniHamed:2000bq}
\bibinfo{author}{\bibfnamefont{N.}~\bibnamefont{Arkani-Hamed}},
  \bibinfo{author}{\bibfnamefont{L.~J.} \bibnamefont{Hall}},
  \bibinfo{author}{\bibfnamefont{H.}~\bibnamefont{Murayama}},
  \bibinfo{author}{\bibfnamefont{D.}~\bibnamefont{Tucker-Smith}},
  \bibnamefont{and} \bibinfo{author}{\bibfnamefont{N.}~\bibnamefont{Weiner}},
  \bibinfo{journal}{Phys.Rev.} \textbf{\bibinfo{volume}{D64}},
  \bibinfo{pages}{115011} (\bibinfo{year}{2001}), \eprint{hep-ph/0006312}.

\bibitem[{\citenamefont{Chatrchyan et~al.}(2012{\natexlab{d}})}]{CMSSUS12011}
\bibinfo{author}{\bibfnamefont{S.}~\bibnamefont{Chatrchyan}}
  \bibnamefont{et~al.} (\bibinfo{collaboration}{CMS Collaboration})
  (\bibinfo{year}{2012}{\natexlab{d}}), \eprint{1207.1898}.

\bibitem[{\citenamefont{Aad et~al.}(2012{\natexlab{f}})}]{Aad:2011cw}
\bibinfo{author}{\bibfnamefont{G.}~\bibnamefont{Aad}} \bibnamefont{et~al.}
  (\bibinfo{collaboration}{ATLAS Collaboration}),
  \bibinfo{journal}{Phys.Rev.Lett.} \textbf{\bibinfo{volume}{108}},
  \bibinfo{pages}{181802} (\bibinfo{year}{2012}{\natexlab{f}}),
  \eprint{1112.3832}.

\bibitem[{\citenamefont{Polesello and Tovey}(2010)}]{Polesello:2009rn}
\bibinfo{author}{\bibfnamefont{G.}~\bibnamefont{Polesello}} \bibnamefont{and}
  \bibinfo{author}{\bibfnamefont{D.~R.} \bibnamefont{Tovey}},
  \bibinfo{journal}{JHEP} \textbf{\bibinfo{volume}{1003}}, \bibinfo{pages}{030}
  (\bibinfo{year}{2010}), \eprint{0910.0174}.

\bibitem[{CMS(2011)}]{CMS-EXO-11-060}
\bibinfo{type}{Tech. Rep.} \bibinfo{number}{CMS-EXO-11-060},
  \bibinfo{institution}{CERN}, \bibinfo{address}{Geneva}
  (\bibinfo{year}{2011}).

\bibitem[{\citenamefont{Ruderman et~al.}(2012)\citenamefont{Ruderman, Slatyer,
  and Weiner}}]{crpv}
\bibinfo{author}{\bibfnamefont{J.~T.} \bibnamefont{Ruderman}},
  \bibinfo{author}{\bibfnamefont{T.~R.} \bibnamefont{Slatyer}},
  \bibnamefont{and} \bibinfo{author}{\bibfnamefont{N.}~\bibnamefont{Weiner}}
  (\bibinfo{year}{2012}).

\bibitem[{ATL(2012{\natexlab{f}})}]{ATLAS-CONF-2012-058}
\bibinfo{type}{Tech. Rep.} \bibinfo{number}{ATLAS-CONF-2012-058},
  \bibinfo{institution}{CERN}, \bibinfo{address}{Geneva}
  (\bibinfo{year}{2012}{\natexlab{f}}).

\bibitem[{ATL(2012{\natexlab{g}})}]{ATLAS-CONF-2012-037}
\bibinfo{type}{Tech. Rep.} \bibinfo{number}{ATLAS-CONF-2012-037},
  \bibinfo{institution}{CERN}, \bibinfo{address}{Geneva}
  (\bibinfo{year}{2012}{\natexlab{g}}).

\bibitem[{\citenamefont{Alves et~al.}(2012)\citenamefont{Alves, Fox, and
  Weiner}}]{SHsignals}
\bibinfo{author}{\bibfnamefont{D.~S.~M.} \bibnamefont{Alves}},
  \bibinfo{author}{\bibfnamefont{P.~J.} \bibnamefont{Fox}}, \bibnamefont{and}
  \bibinfo{author}{\bibfnamefont{N.}~\bibnamefont{Weiner}},
  \emph{\bibinfo{title}{Enhanced higgs signals from a type i 2hdm and sister
  higgs}} (\bibinfo{year}{2012}).

\end{thebibliography}
\bibliographystyle{apsrev}

\end{document}